\documentclass[twocolumn]{aastex631}

\usepackage{amsmath}
\usepackage{multirow}
\usepackage{gensymb}
\usepackage{siunitx}
\begin{document}

\title{Evidence of Water Vapor in the Atmosphere of a Metal-Rich Hot Saturn with High-Resolution Transmission Spectroscopy}

\author[0009-0002-6791-3982]{Sayyed A. Rafi}
\affiliation{Department of Astronomy, University of Tokyo, 7-3-1 Hongo, Bunkyo-ku, Tokyo 113-0033, Japan}
\email{salirafi8@gmail.com}
\email{salirafi@g.ecc.u-tokyo.ac.jp}

\author[0000-0003-4698-6285]{Stevanus K. Nugroho}
\affiliation{Astrobiology Center, 2-21-1 Osawa, Mitaka, Tokyo 181-8588, Japan}
\affiliation{National Astronomical Observatory of Japan, 2-21-1 Osawa, Mitaka, Tokyo 181-8588, Japan}

\author[0000-0002-6510-0681]{Motohide Tamura}
\affiliation{Department of Astronomy, University of Tokyo, 7-3-1 Hongo, Bunkyo-ku, Tokyo 113-0033, Japan}
\affiliation{Astrobiology Center, 2-21-1 Osawa, Mitaka, Tokyo 181-8588, Japan}
\affiliation{National Astronomical Observatory of Japan, 2-21-1 Osawa, Mitaka, Tokyo 181-8588, Japan}

\author[0000-0001-8419-8760]{Lisa Nortmann}
\affiliation{Institut für Astrophysik und Geophysik, Georg-August-Universität, D-37077 Göttingen, Germany}

\author[0000-0002-0516-7956]{Alejandro Sánchez-López}
\affiliation{Instituto de Astrofísica de Andalucía (IAA-CSIC), Glorieta de la Astronomía s/n, 18008 Granada, Spain}

\begin{abstract}

Transmission spectroscopy presents one of the most successful approaches for investigating the atmospheres of exoplanets. We analyzed the near-infrared high-resolution transmission spectrum of a hot Saturn, HD 149026 b, taken using CARMENES spectrograph ($\mathcal{R}\sim80,400$). We found evidence of H$_2$O at an S/N of $\sim$4.8. We also performed grid search using a Bayesian framework and constrained the orbital velocity $K_\mathrm{p}$ and rest velocity $V_{\mathrm{rest}}$ to $158.17^{+8.31}_{-7.90}$ $\mathrm{km\ s}^{-1}$ and $2.57^{+0.54}_{-0.57}$ $\mathrm{km\ s}^{-1}$, respectively. Whilst the retrieved $K_\mathrm{p}$ value is consistent with theoretical prediction, the retrieved $V_{\mathrm{rest}}$ value is highly red-shifted ($>$3-$\sigma$). This might be an indication of either anomalous atmospheric dynamics at play or an orbit with non-zero eccentricity. Additionally, we searched for HCN but no successful detection has been made possibly due to the relatively low S/N dataset. The detection of H$_2$O and subsequent abundance retrieval, coupled with analysis of other species such as CO at the $K$-band, for example, might help us to get some information about the atmospheric C/O ratio and metallicity, which in turn could give us some insight into the planet formation scenario.

\end{abstract}


\section{Introduction}\label{chap:intro}

In 2000, \cite{Seager_2000} theorized that during a planetary transit the planet's atmosphere can be observed via transmission spectroscopy. Just two years later, the very first detection of an exoplanet atmosphere was made by \cite{Charbonneau_2002}, which detected sodium doublet absorption at 589.3 nm in the atmosphere of HD 209568 b using HST/STIS transmission data. Since then, great effort has been put into characterizing hot Jupiter atmospheres using low-resolution, both via transit and emission spectroscopy (e.g., \citealp{Charbonneau_2002, Deming_2006, Tinetti_2007, Machalek_2009}). 

Later, using Cryogenic Infrared Échelle Spectrograph with $\mathcal{R}\sim100,000$, mounted on the Very Large Telescope, carbon monoxide (CO) was detected by \cite{Snellen_2010} in the atmosphere of HD 209458 b while it transits the star. The authors also constrained its orbital velocity, allowing them to derive the absolute planetary mass. Interestingly, the detected CO signal is blue-shifted by $2\pm1$ $\mathrm{km\ s}^{-1}$ relative to the expected location, indicating possible high-altitude winds in the atmosphere. This became the first direct detection of a planetary-scale wind in the atmosphere of an exoplanet. Since then, the role of high-resolution spectroscopy in exoplanet atmospheric science underwent significant advancements and remains to be a powerful technique even in the era of James Webb Space Telescope (JWST). There are many hot Jupiters of which atmospheres have been characterized using this method, including the detection of atomic \citep[e.g.][]{Hoeijmakers_2019, Borsa_2021, Kesseli_2022, Stangret_2022} and molecular species such as H$_2$O \citep[e.g.][]{Alonso_Floriano_2019, Sanchez-Lopez_2019, Giacobbe_2021, Webb_2022}, HCN \citep[e.g.][]{Cabot_2018, Giacobbe_2021, Sanchez-Lopez_2022}, CH$_4$ \citep[e.g.][]{Guilluy_2019, Giacobbe_2021}, CO \citep[e.g.][]{Snellen_2010, Giacobbe_2021, Ramkumar2023}, OH \citep{Nugroho_2021, Landman_2021, Brogi_2023, Wright2023}, and even TiO and VO \citep{Nugroho_2017, Cont2021, Prinoth_2022, Pelletier_2023}. 

Recently, interests begin to shift to atmospheric retrieval (e.g., \citealp{Brogi_2019, Gibson_2020, Gibson_2022, Line_2021, Pelletier_2021, Kawahara_2022, Brogi_2023, Yan_2023, Blain_2024}) and advanced 3D analyses (e.g., \citealp{Flowers_2019, Wardenier_2021, vanSluijs_2022}). Constraining the chemical abundances of the atmosphere through atmospheric retrieval could provide us with various useful information about the planet, including its planetary formation history. For hot gas giants, one way to do it is to use the C/O ratio as a tracer to the formation and migration scenario (e.g., \citealp{Oberg_2011, Madhusudhan_2014, Mordasini_2016, Brewer_2017, Cridland_2019, Schneider_2021}). Tracer species with sufficiently high abundances (at a given atmospheric condition) such as CO, H$_2$O, HCN, CH$_4$, and CO$_2$ are often used to constrain this parameter. In solar composition for planet hotter than $\sim$ 1800 K, for example, it is reasonable to use both H$_2$O and CO as proxies to the C/O ratio, since we would expect that C and O are abundant enough to be representative of all other metals \citep{Asplund_2009} and that H$_2$O and CO are the dominant O- and C-bearing trace species in the atmosphere \citep{Madhusudhan_2012, Moses_2013, Molliere_2015}. While for super-solar composition (roughly C/O $>$ 1 for equilibrium temperature $>$ 1000 K), HCN and CH$_4$ will instead become the dominant C-bearing species. Hence, detecting and subsequently constraining the abundance of the dominating species is key to infer the planetary atmospheric C/O ratio, and eventually get a glimpse of the planetary formation and migration history.

One of the most intriguing hot gas giants in planetary formation context may be HD 149026 b, a hot Saturn orbiting a metal-rich evolved star ($[\mathrm{Fe/H}]=0.36$; G0 IV) within a 2.8-day of orbital period \citep{Bonomo_2017}, making the atmosphere heated up such that the equilibrium temperature of the planet reaches $\sim1700$ K (following formula in \cite{Guillot_2010}, assuming planetary zero-albedo and complete heat redistribution from the day-side to the night-side). Table \ref{tab:stellar_planet_param} shows the orbital and physical parameters of the planet and the host star together with the references. \cite{Sato_2005} found that the planet has an anomalously large core ($43.6-89.3$ $M_\oplus$) that challenges the existing planet formation theories, including gravitational instability and core accretion. This was then supported by later studies that eventually put a maximum core mass up to 110 $M_\oplus$ (e.g., \citealp{Fortney_2006, Broeg_2007, Burrows_2007}). Indeed, the actual formation scenario for this planet is still poorly understood. \cite{Ikoma_2006} proposed that a catastrophic collision involving two or more giant planets or an unusually high supply of planetesimals by gravitational perturbations from other planets is necessary to form such a heavy core (which indicates that HD 149026 might be a multi-planetary system). \cite{Broeg_2007} proposed an unusual scenario where HD 149026 b could form in-situ, assuming that the planetesimal flux was anomalously high (10$^{-2}$ M$_{\oplus}$ yr$^{-1}$). \cite{Dodson_Robinson_2009} then showed that the standard core accretion scenario with elevated solid surface density and large initial orbit would be sufficient to form the planet. 

Several recent observational works may provide some insights into the problem. Using Subaru/HDS high-resolution optical data of HD 149026 b, \cite{Ishizuka_2021} detected Ti but could not detect TiO, which could be explained by the planet having a super-solar C/O ratio (as the C/O increases, the abundance of TiO significantly decreases). They showed that their result is also consistent with C/O $\gtrsim$ 1. Recently, using JWST/NIRCam thermal emission data, \cite{Bean_2023} detected H$_2$O and CO$_2$ in its atmosphere and constrained the metallicity and C/O ratio to be super-solar with [M/H] = 2.09$^{+0.35}_{-0.32}$ and C/O = 0.84$^{+0.03}_{-0.03}$. This indicates that either HD 149026 b was formed via core accretion and then accreted its atmosphere exterior to H$_2$O snowline where most of the oxygen locked in H$_2$O are present in solids and underwent a disk-free migration or it was formed in an already carbon-rich disk \citep{Matsuo_2007, Oberg_2011, Madhusudhan_2014, Cridland_2019}. The retrieved C/O value from \cite{Bean_2023} may also imply that the atmosphere is roughly in the oxygen- to carbon-rich boundary (assuming equilibrium temperature; \citealp{Molliere_2015}). This may suggest that H$_2$O, HCN, and CH$_4$ may be important in building the C/O ratio. 

\begin{table}[t]
\centering
\begin{tabular}{cc}
\hline
\textbf{Stellar parameters} &  \textbf{Value}\\
\hline
Name &  HD 149026 \\
Mass ($M_\odot$) &  1.34 $\pm$ 0.02$^\mathrm{a}$ \\
Radius ($R_\odot$) &  1.41 $\pm$ 0.03$^\mathrm{b}$ \\
Effective temperature (K) &  6179 $\pm$ 15$^\mathrm{b}$ \\
Metallicity (dex) &  0.36 $\pm$ 0.08$^\mathrm{a}$ \\
Distance (pc) & 76.218 $\pm$ 0.0941$^\mathrm{c}$ \\
$V_\mathrm{mag}$ &  8.15$^\mathrm{d}$ \\
$H_\mathrm{mag}$ &  6.899$^\mathrm{d}$ \\
\hline
\textbf{Planetary parameters} &  \textbf{Value} \\
\hline
Name &  HD 149026 b \\
Mass ($M_J$) &  0.322$^{+0.014}_{-0.012}$$^\mathrm{a}$ \\
Radius ($R_J$) &  0.811$^{+0.029}_{-0.027}$$^\mathrm{a}$ \\
Semi-major axis (AU) &  0.04364 $\pm$ 0.00022$^\mathrm{a}$ \\
Inclination ($\degree$) &  84.50$^{+0.60}_{-0.52}$$^\mathrm{a}$ \\
Period (days) &  2.87588874 $\pm$ 0.00000059$^\mathrm{e}$ \\
Transit midpoint (BJD$_{\mathrm{UTC}}$) &  2454456.78760 $\pm$ 0.00016$^\mathrm{e}$ \\
Equilibrium temperature (K) & 1693 \\
Eccentricity & 0.0 \\
\hline
\end{tabular}

\caption{The adopted stellar and planetary parameters in this work. References are also given. The equilibrium temperature is calculated assuming perfect day-to-night recirculation and zero albedo \citep{Guillot_2010}.\\
\textbf{Note.} $^\mathrm{(a)}$ \cite{Bonomo_2017}, $^\mathrm{(b)}$ \cite{Stassun_2017}, $^\mathrm{(c)}$ \cite{Gaia_2020}, $^\mathrm{(d)}$ SIMBAD Database, $^\mathrm{(e)}$ \cite{Zhang_2018}}
\label{tab:stellar_planet_param}
\end{table}

In this work, we searched for H$_2$O and HCN in the atmosphere of the dense hot Saturn HD 149026 b using high-resolution cross-correlation spectroscopy (HRCCS; \citealp{Brogi_2019}). H$_2$O is an important molecule for a wide regime of temperature of interest and for C/O less than unity, as this molecule is the most abundant O-bearing species together with CO. Hence, detecting H$_2$O is key as the first step to understand the planet's atmosphere. This is also to confirm the finding of H$_2$O from \cite{Bean_2023} with different type of both data and analysis technique as well as qualitatively assess whether the planet has a super-solar C/O ratio atmosphere (of which HCN might be present). 

We organized this paper as follows. Section \ref{chap:obs} describes the observation data that we used and outlined the data reduction procedures until the clean spectrum (stellar and telluric lines are removed) is obtained. Section \ref{chap:model} describes the models used in this work and outlined our procedures to pre-process the model to mimic the altered planet signal after applying SysRem. Section \ref{chap:searching} describes the cross-correlation method to detect planet signals. We then assessed the detection significance using two methods, S/N map and Welch-t test, and constrained the planetary orbital and systemic velocity in Section \ref{chap:results}. Finally, we discuss our results in Section \ref{chap:discussion} and conclude our work in Section \ref{chap:summary}.

\vfill\null
\section{Observations and Data Reduction} \label{chap:obs}

\subsection{Observations}

HD 149026 b was observed with CARMENES installed in 3.5-m Calar-Alto Observatory for one night during its transit on 2019 April 12 (PI: Nortmann, L., PID: F19-2.0-023), obtaining 72 exposures\footnote{\href{http://caha.sdc.cab.inta-csic.es/calto/jsp/searchform.jsp}{http://caha.sdc.cab.inta-csic.es/calto/jsp/searchform.jsp}}. CARMENES has two separated channels, each for visual (VIS) and near-infrared (NIR) wavelengths. We only used the NIR channel in this work, which spans from 0.96 to 1.71 $\mu$m (YJH band) consisting of 28 echelle orders with a resolution of $\mathcal{R}\sim80,400$. The light is collected by two 2kx2k Hawaii-2RG detectors which are physically separated, resulting in a gap seen in every spectral order. The raw 2-D spectra were reduced and converted to 1-D spectra by CARMENES automatic reduction pipeline, \texttt{caracal v2.10} \citep{Caballero_2016}, which performs the standard reduction procedure: dark and bias subtraction, non-linearity curves correction, order tracing, optimal extraction, and wavelength calibration in vacuum (which is also used throughout this work), as well as providing the barycentric radial velocity (RV) correction. The reduced spectra are given in the telluric rest frame. Each CARMENES channel is fed by two fibers, fiber A and fiber B, of which the former is targeted to the star, obtaining the science spectrum, while the latter can be used to obtain either simultaneous wavelength calibration with the Fabry-Perot etalon or the sky spectrum used to identify the sky emission lines. Although fiber B was used for obtaining the sky spectrum in this observation, we did not use it in our analysis. Instead, we removed the sky emission lines by masking them based on the two near-infrared sky emission atlas from \cite{Rousselot2000} and \cite{Oliva_2015} (see Section \ref{chap:data_reduc}). 

\subsection{Data reduction} \label{chap:data_reduc}

\begin{figure}[t!]
\centering
\includegraphics[width=\linewidth]{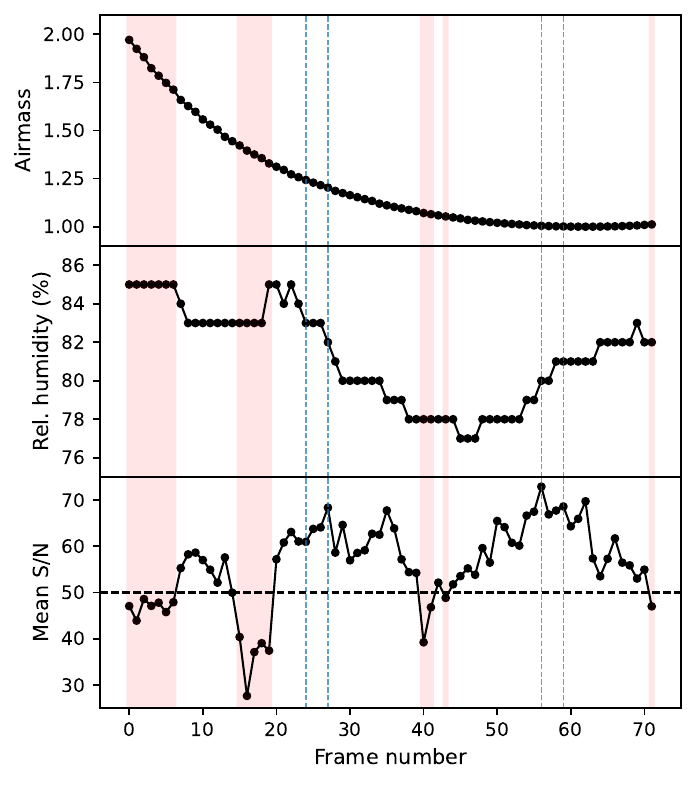}
\caption{\textbf{Top}: Airmass throughout the observation. \textbf{Middle}: Relative humidity. \textbf{Bottom}: Average S/N over orders of each exposure. Regions highlighted in light red color are exposures excluded from the analysis. The horizontal black dashed line corresponds to S/N = 50. One exposure has an average S/N of $\sim$49.9 (frame \#14), hence, given the few number of frames, we decided to keep this one for the analysis. The vertical dashed blue lines mark the first to fourth transit contact starting from the left.}
\label{fig:night_condition}
\end{figure}

\begin{figure}[t!]
\centering
\includegraphics[width=\linewidth]{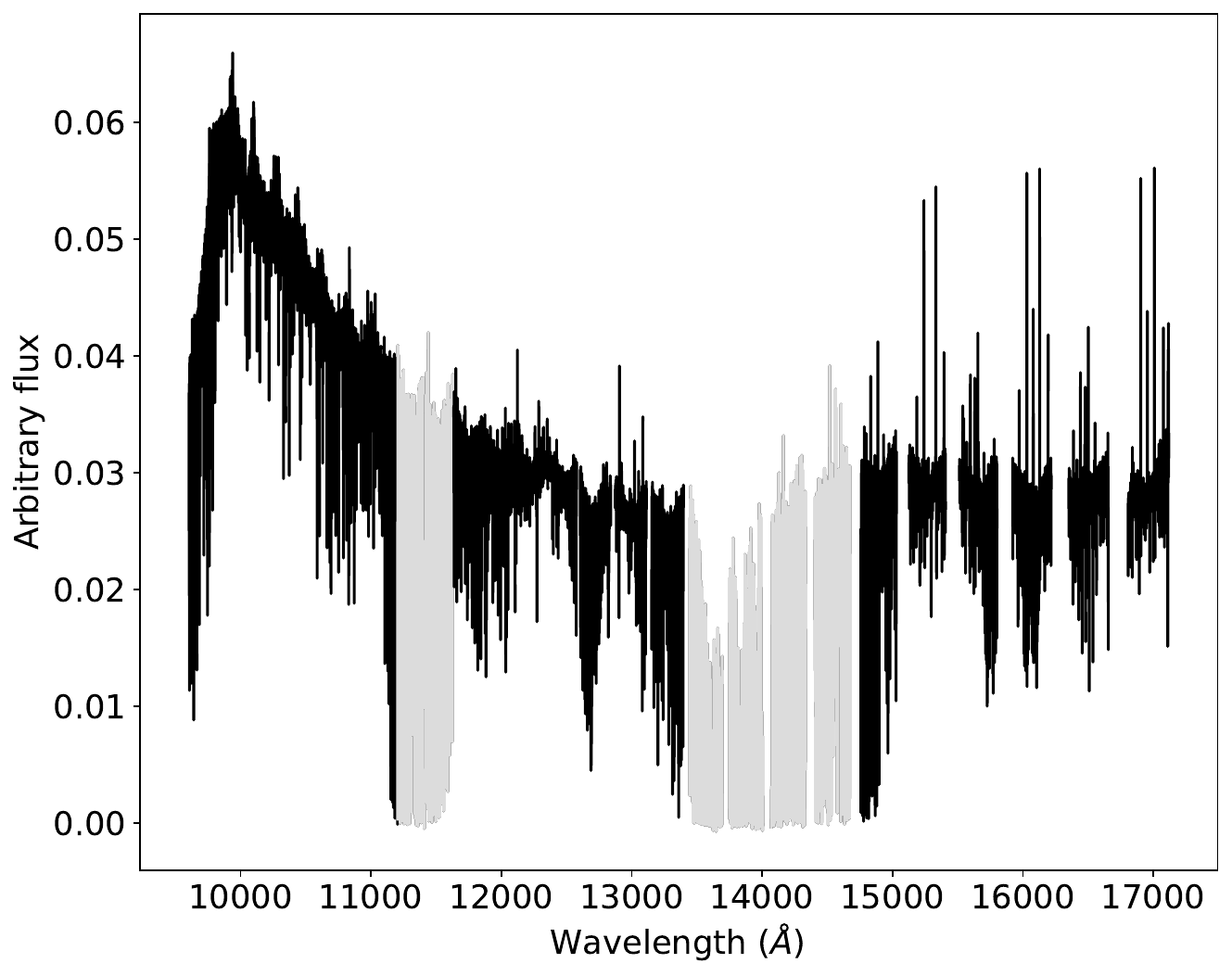}
\caption{The full \texttt{caracal}-reduced 1-D spectrum for frame \#56 (highest S/N frame). Regions highlighted with light gray color are orders that are removed from the analysis due to their low S/N. These orders are the strongest bands of telluric water absorption. However, as can be seen in, some of the strong telluric bands still remain, especially in the reddest part of order 55 ($\sim$1.1 $\mu m$) and bluest part of order 41 ($\sim$1.48 $\mu m$). This will be masked as explained in Section \ref{chap:data_reduc}. The V-like shapes seen within each order are likely instrumental effects or artifacts from the reduction pipeline \citep{Alonso_Floriano_2019}.}
\label{fig:masked_orders}
\end{figure}

Figure \ref{fig:night_condition} shows the night condition during the observation along with the obtained S/N. The high level of relative humidity at the time of observation ($\sim$81\% on average) resulted in the relatively poor quality of the data, with an average S/N per pixel over all exposures is only $\sim$56 (far below the expected S/N from a bright star like HD 149026 with $H_{\mathrm{mag}}=6.89$; S/N $>150$). The airmass itself varied from $\sim2$ to $\sim1$ from the beginning to the end of the observation. We decided to exclude the first seven exposures with the highest airmass and also those with significant drops in S/N i.e., exposures with S/N less than 50 (Figure \ref{fig:night_condition}). In addition, spectral orders corresponding to the strongest absorption bands of telluric water (order 54-53 and 45-42) also have very low S/N, hence they are excluded from the remainder of the analysis (Figure \ref{fig:masked_orders}). In the end, we have 56 frames (of which 32 of them belong to the in-transit phase) with each having 22 useful orders to be further analyzed. 

Additional reduction procedures were performed as follows. After masking negative values, we removed the instrumental throughput variations across wavelengths by normalizing each spectrum over the pseudo-continuum of a reference spectrum (fitted using \texttt{continuum} task in \texttt{IRAF}), which is chosen as the spectrum with the highest S/N (frame \#56; Figure \ref{fig:night_condition}). To remove flux variations between frames (top panel of Figure \ref{fig:spec_anal}), we normalized each spectrum by dividing over its respective median value. 

As for sky emission lines which are prominent in near-infrared, especially for redder wavelengths as seen in Figure \ref{fig:masked_orders}, they were identified and subsequently masked using the near-infrared emission lines atlas from \cite{Rousselot2000} and \cite{Oliva_2015} which mostly contain OH emissions as well as several non-OH and unidentified lines. We noticed an unusual 'bump' at the bluest part of order 61, 60, and 59 (with median wavelength of 10,023 \si{\angstrom}, 10,190 \si{\angstrom}, and 10,362 \si{\angstrom} respectively) which lies at the same pixel and might arise from bad regions in the detector, not astrophysical, hence we masked them from the analysis. In addition, we also masked 40 pixels ($\sim1\%$ from the total number of pixels per order) from the edges of each order to account for edge-effect (where edges of each order have the lowest S/N). Following \cite{Gibson_2020}, we removed outliers in each spectral order by first constructing a 2-D matrix with time (frames) as columns and wavelength as rows (as shown in Figure \ref{fig:spec_anal}). We then computed a simple model by computing the outer product of the median of each spectrum (along the wavelength axis) and the median of the time-series (along the time axis) and divide it by the median of the data 2-D map for normalization. We subtracted the data 2-D map by the model and fit a 10-th-order polynomial to the residual. We then masked any 5-$\sigma$ deviations from the fitted continuum with $\sigma$ taken as the standard deviation of the residual. We also masked pixels associated with cosmic rays which are not perfectly removed by \texttt{caracal} by performing 5-$\sigma$ clipping for each wavelength channel. Finally, we masked the strongest telluric lines which cores lie below 30\% of the continuum levels. We performed this by identifying pixels that lie below the threshold level and subsequently masked the surrounding pixels (the line wings) that fall between a defined range expressed by
\begin{equation}
    |\Delta\lambda|<\frac{\eta}{2}\frac{\lambda_{\mathrm{tel}}}{R}
\end{equation}
where $\lambda_{\mathrm{tel}}$ is the wavelength that corresponds to the core of the strong telluric line, $R$ is the data resolution, $\eta$ is a multiplication factor, and $\Delta\lambda=\lambda-\lambda_{\mathrm{tel}}$. Discussion about the effect of $\eta$ on the resulting detection significance will be addressed further in Section \ref{chap:param_eta}, but for now, we will just state that $\eta=18$ provided us with the best detection significance, assuming the threshold level is 30\% the continuum. In total, for every exposure, we masked 38\% of the data points.

We follow \cite{Nugroho_2020a} to correct for the residual blaze function not perfectly removed by \texttt{caracal}. This is also to make sure that the spectra are within a similar continuum level. The procedure performed is as follows. For each 2-D matrix of every spectral order, we masked any strong telluric absorption not removed by the previous masking procedure and divided the matrix by the median of the time series to remove the dominant spectral features and smooth the spectrum. Then, for each frame of the residual, a median and a Gaussian filter\footnote{\href{https://docs.scipy.org/doc/scipy/reference/generated/scipy.ndimage.gaussian_filter1-D.html}{scipy.ndimage.gaussian\_filter1-D}} are passed to produce the smoothed pseudo-continuum. Note that we have to choose a wide enough bin size and standard deviation for the median and Gaussian filters, respectively, in order to not alter the underlying planet signal in the spectrum \citep{Gibson_2018, Gibson_2022}. We chose 501 and 100 pixels for the median and Gaussian filters, respectively. Each spectrum is then divided over the smoothed pseudo-continuum to correct for any continuum variations in between exposures.

As a reference, we also investigated the stability of the wavelength solution given by \texttt{caracal}. We cross-correlated spectral orders that have many strong telluric lines with a model telluric transmission spectrum calculated using ESO Sky Model Calculator\footnote{\href{https://www.eso.org/observing/etc/bin/gen/form?INS.MODE=swspectr+INS.NAME=SKYCALC}{https://www.eso.org/observing/etc/doc/skycalc/helpskycalc.html}} across a grid of RV values. We then used a likelihood-based approach defined in \cite{Gibson_2020} which maps the resulting cross-correlation values to likelihood to derive the best-fit RV (see Section \ref{chap:likelihood}). We found that the shifts for all exposures are at sub-pixel level with a mean of $\sim$0.057 $\mathrm{km\ s}^{-1}$ which is much smaller than the data FWHM ($\sim$3.73 $\mathrm{km\ s}^{-1}$) and thus no correction is required.

\vfill\null
\subsection{Stellar and telluric lines removal} \label{chap:SYSREM}

\begin{figure}[t]
\centering
\includegraphics[width=\linewidth]{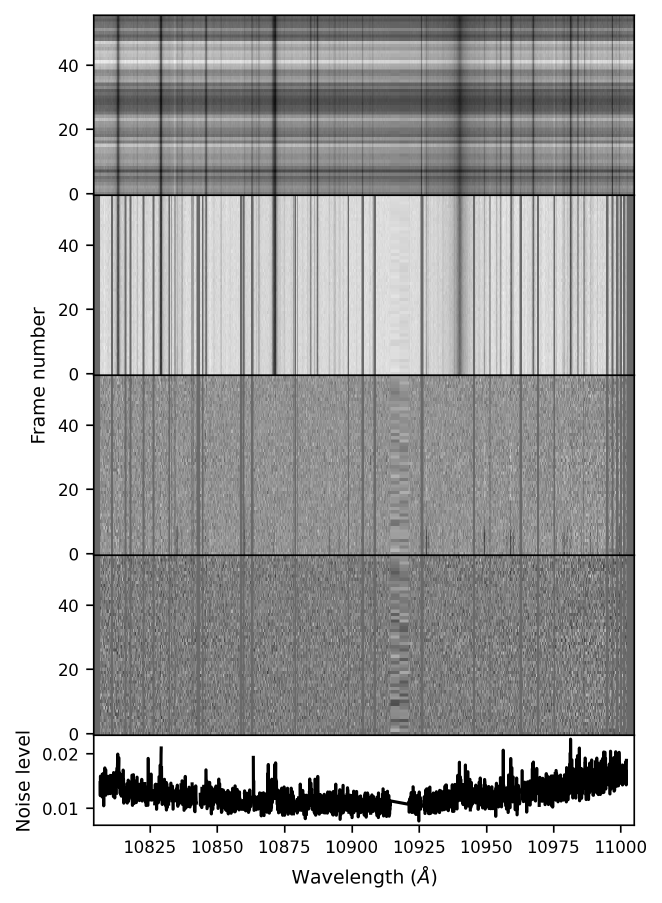}
\caption{Spectra of order 57 throughout the analysis outlined in Section \ref{chap:obs} in a 2-D matrix (time x wavelength). The first panel shows the output spectra of \texttt{caracal}. The variation of flux between frames is caused by the difference in the S/N due to different observational conditions, instrumental effects, etc. The second panel shows the spectra after normalization, outliers masking, and continuum-level correction. Gray regions in the color map are masked pixels. The third panel shows the spectra after median normalization. The fourth panel shows the spectra after applying SysRem for one iteration. The bottom panel shows the standard deviation from each wavelength channel in the residual.}
\label{fig:spec_anal}
\end{figure}

To remove the dominating stellar and telluric lines, we used a detrending algorithm called SysRem \citep{Tamuz_2005} which has been widely used in literature \citep[e.g.][]{Birkby_2017, Nugroho_2017, Nugroho_2020a, Alonso_Floriano_2019, Gibson_2018, Gibson_2020, Ishizuka_2021, Herman_2022, Yan_2023}. The algorithm is originally developed to remove atmospheric extinction and any linear systematic trends that might be present in a set of photometric light curves. In the high-resolution spectroscopy analysis, SysRem treats each wavelength bin in the 2-D map (Figure \ref{fig:spec_anal}) as a light curve, then fits and removes the quasi-static stellar and telluric lines (and any other linear systematics). One advantage of using SysRem compared to other detrending algorithms is that it takes into account different uncertainty pixel-by-pixel, hence, assuming that we have a proper estimate for the uncertainty, an accurate fit for the systematics can be obtained.

We followed the procedure outlined in \cite{Gibson_2022} by first dividing the spectrum by the median of the time series to properly normalize the data (this would remove the stellar lines but not the telluric lines which vary in flux during observations due to changing airmass and precipitable water vapor; third panel of Figure \ref{fig:spec_anal}). We then estimated the uncertainty of each pixel in each spectral order by computing the outer product of the standard deviation of the spectrum (along the wavelength axis) and the standard deviation of the time series (along the time axis), divided by the standard deviation of the 2-D map \citep{Nugroho_2020a, Ridden-Harper_2023}. We passed the resulting spectra and estimated uncertainties directly into the SysRem algorithm and performed 10 SysRem iterations. The best-fit SysRem model of each iteration was then subtracted from the data. Finally, we masked the remaining outliers for each SysRem iteration in the same manner as has been described in Section \ref{chap:data_reduc}. The bottom panel of Figure \ref{fig:spec_anal} shows the residuals after one SysRem iteration. All of the dominant systematics, such as the residual telluric lines, are now removed, at least visually. This is the final spectra that will be cross-correlated with the planetary spectrum model.

Performing only one SysRem iteration might not be enough to completely remove the telluric lines. Moreover, the optimum number of iterations for each order could be different as they have different levels of telluric contamination. One important thing to note is that SysRem is also slowly removing the planet signal from the beginning of the iteration. Therefore, getting the best detection significance from the planet signal is a matter of balancing these two considerations: choosing the iteration at which the systematics are removed completely while keeping the planetary signal intact. Several studies instead used the variance of the SysRem residuals of each spectral order to determine the best SysRem iteration, which is the iteration where the variance starts to plateau \citep[e.g.][]{Deibert_2021, Herman_2020, Herman_2022, Ridden-Harper_2023}. However, to avoid bias in our results by adopting different SysRem iterations between spectral orders \citep{Cabot_2018}, we used the same number of iterations for all orders \citep[e.g.][]{Nugroho_2020a, Nugroho_2020b} and adopted the iteration at which the significance is the highest for subsequent analyses unless stated otherwise. We note, however, that this choice means that some of the telluric contamination would not be perfectly removed in some orders. 

\begin{figure}
\centering
\includegraphics[width=\linewidth]{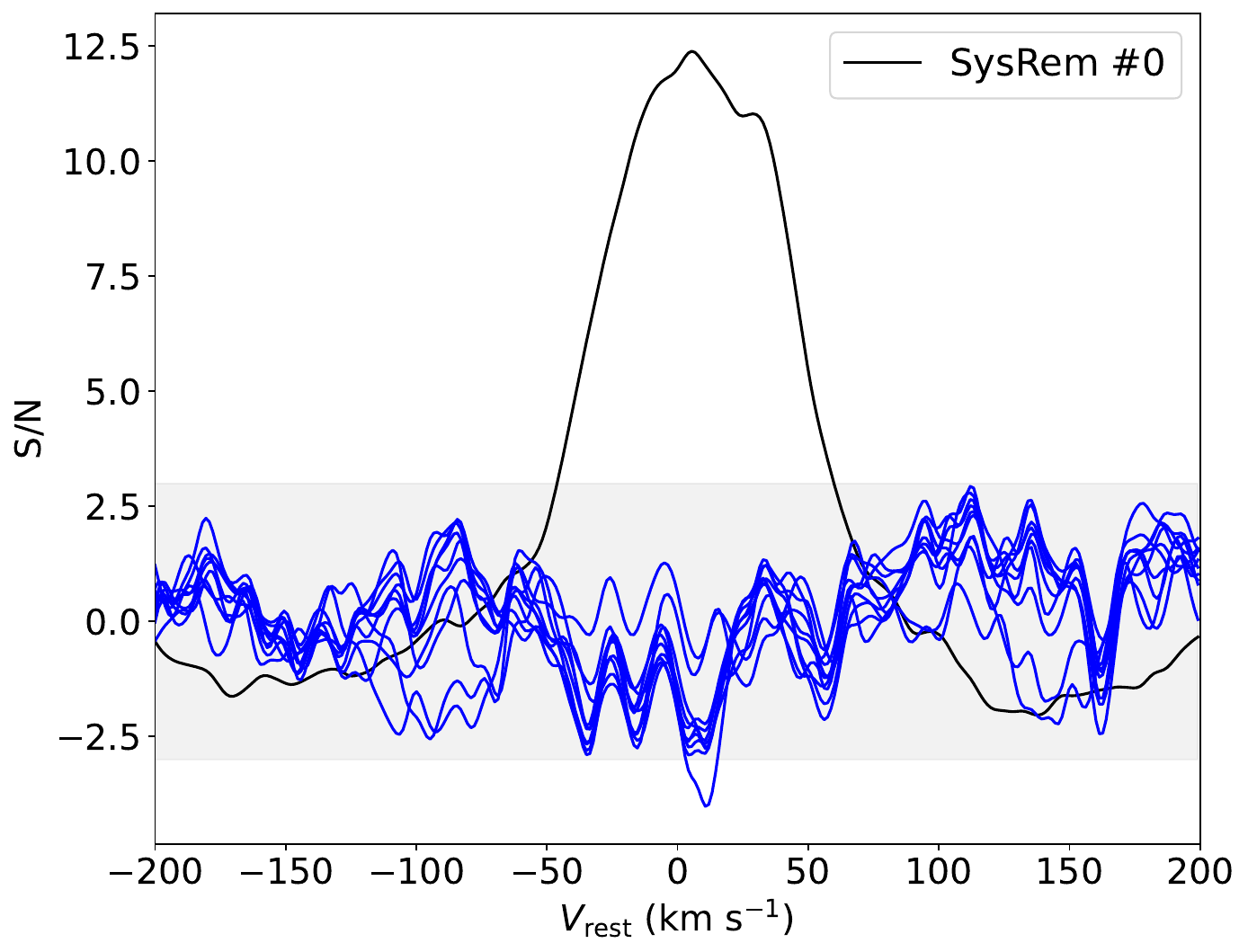}
\caption{Slices of telluric CCF (cross-correlation between the SysRem residual spectra and telluric model) at the expected planet location, $K_\mathrm{p} =$ 164.27 $\mathrm{km\ s}^{-1}$ (see Section \ref{chap:searching}), for different SysRem iterations. Gray area highlights regions within S/N $\pm$ 3. SysRem \#0 means before applying SysRem. The blue curves are all CCF values after applying SysRem (from SysRem \#1 to \#10).}
\label{fig:telluric_ccf}
\end{figure}

In order to check whether the telluric residual is significant or not after applying SysRem, we cross-correlated the residuals with a telluric model (used to calculate the telluric RV; Section \ref{chap:data_reduc}), constructed the telluric S/N map (see Section \ref{chap:searching} for details about constructing the S/N map) and then took a slice of the map at the expected planet location. The result is shown in Figure \ref{fig:telluric_ccf}. We can see that, by only dividing the data with median spectrum (i.e. before applying SysRem, SysRem \#0), the telluric signal still dominates the map, seen from the low-noise pattern and relatively high S/N. Once we applied SysRem, the telluric signal completely disappears, even only after performing one iteration. The CCF then shows similar curve trend with increasing iterations. Thus, this test suggests that telluric residuals may not be able to explain our signal.

\section{Planetary Spectrum Modeling}\label{chap:model}

We generated the planetary spectrum models using the publicly available radiative transfer code, \texttt{petitRADTRANS}\footnote{\href{https://petitradtrans.readthedocs.io/en/latest/index.html}{https://petitradtrans.readthedocs.io/en/latest/index.html}} \citep{Molliere_2019}. We assumed a 1-D plane parallel cloud-free atmosphere with an isothermal temperature-pressure (T-P) profile. Whilst the assumed isothermal profile is unlikely to be true in most cases, it is still a reasonable approximation for a transmission spectrum \citep{Fortney_2005}. We divided the atmosphere into 100 atmospheric layers equally spaced in log pressure from 10$^{-8}$ to 100 bar and set the reference pressure at 10 bar \citep{Fortney_2010,Tinetti_2012}.

CIA from H$_2$-He and H$_2$-H$_2$ collision and Rayleigh scattering from H$_2$ and He are included as the continuum opacity sources. In addition, we also included bound-free and free-free absorption from H$^-$. We included two trace gases in our models: H$_2$O and HCN. For HCN, we used the high-resolution pre-computed opacity grid provided in \texttt{petitRADTRANS} produced using ExoMol database \citep{Barber_2014,Harris_2006}. It is known that different line lists might affect our capability in detecting the signal \citep[e.g.][]{Gandhi_2020, Webb_2020, Nugroho_2021}, hence we wanted to check whether this discrepancy appears in our results and simultaneously assess our detection robustness (Section \ref{chap:results}). Therefore, for H$_2$O, we used pre-computed opacity grids computed using two different line lists: HITEMP2010 \citep{Rothman_2010} and POKAZATEL from Exomol \citep{Polyansky_2018}, where both opacity grids were computed using \texttt{HELIOS-K} \citep{grimm2015,Grimm_2021} at a resolution of 0.005 cm$^{-1}$ with a line wing cutoff of 100 cm$^{-1}$ for a grid of temperature and pressure profile adopted from \texttt{PTgrid\_new.dat} in the \texttt{petitRADTRANS} webpage. 

\begin{figure*}[t]
\centering
\includegraphics[width=0.85\linewidth]{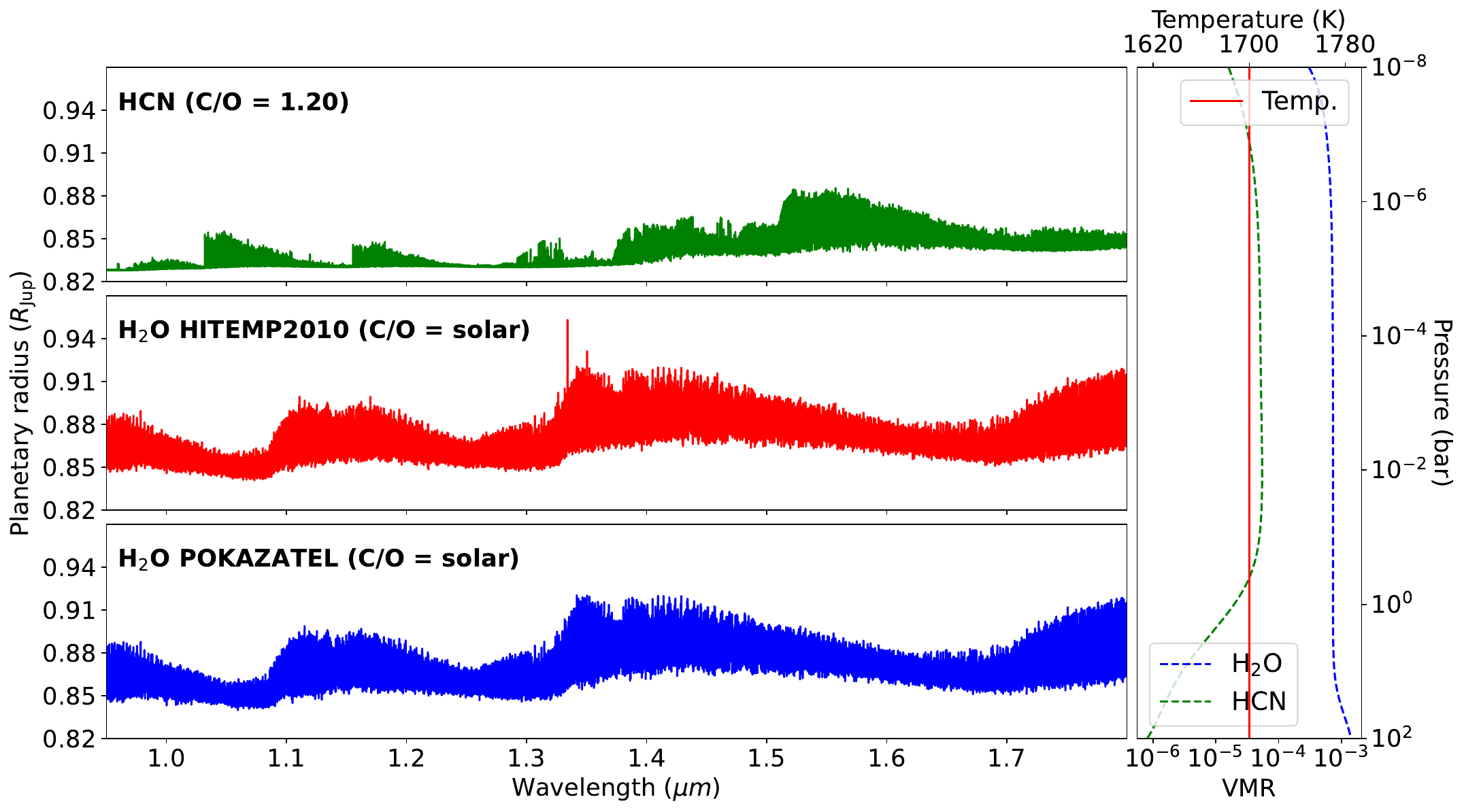}
\caption{\textit{Left panels} show the chemical equilibrium model spectra generated using \texttt{petitRADTRANS} in planetary radius. The models shown are already broadened by FWHM of CARMENES. \textit{Right panel} shows the temperature and chemical equilibrium abundance profiles for H$_2$O (solar C/O) and HCN (C/O = 1.2), computed using FastChem, used to generate the spectra in the left panels.}
\label{fig:models}
\end{figure*}

We described our models as follows. We assumed an isothermal cloud-free atmosphere with a temperature of 1700 K, close to the theoretical equilibrium temperature assuming zero albedo (Table \ref{tab:stellar_planet_param}). We created a single-species model for each of H$_2$O (both HITEMP and POKAZATEL line lists) and HCN. We computed the chemical abundances using \texttt{FastChem} \citep{Stock_2018} assuming chemical equilibrium with solar C/O ratio for H$_2$O and C/O = 1.20 for HCN. The assumption of a super-solar C/O ratio for HCN is based on transmission spectroscopy analysis by \cite{Ishizuka_2021} whose findings are consistent with C/O $\gtrsim$ 1. Consequently, we aim to investigate whether our data can detect both H$_2$O and HCN, as will be discussed in Section \ref{chap:discussion}. We assumed the atmospheric metallicity to be similar with the host star, where [M/H] = 0.36 dex measured relative to solar metallicity with elemental abundances from \cite{Asplund_2009}. In total, we generated three chemical equilibrium models to search for the corresponding species. Figure \ref{fig:models} shows the models together with the chemical equilibrium abundance profiles for H$_2$O and HCN as well as the isothermal T-P profile.

The spectrum models were then convolved to the resolution of CARMENES using a Gaussian kernel. Finally, since the continuum of the planet signal was removed when we performed the procedures outlined in Section \ref{chap:obs}, we normalized the model by subtracting the baseline level computed using a custom maximum high-pass and Gaussian filter with a window and standard deviation of 2000 and 1500 pixels, respectively.

From here, our work used two different approaches: the unprocessed and the processed-model cross-correlation (CC) approach. The latter is based on the fact that a detrending algorithm like SysRem alters the underlying planet signal, as has been discussed in Section \ref{chap:SYSREM}. Therefore, ideally, the model spectrum should also be (pre-)processed to mimic the effect given by SysRem and other analysis procedures that might alter the planet signal. This is especially important whenever retrieval analysis is desired \citep{Gibson_2022}. When the goal is to only search and detect atmospheric species, model pre-processing, in principle, would help to enhance the detection strength but is not necessarily needed. Hence, in this work, we used the unprocessed-model CC approach to search for our species of interest and the processed-model CC approach for subsequent retrieval of the orbital parameters. We used the method outlined in \cite{Gibson_2022} to pre-process the model where we constructed the basis models for each of the 2-D map \citep[equation 7 in][]{Gibson_2022} from the basis vectors of the fitted SysRem models i.e., the $a$ coefficient in \cite{Tamuz_2005}, then subtracted out the basis models from our modeled transmission spectra. We note that we did not perform the continuum subtraction for the pre-processed models, as the procedure will automatically remove it.

\section{Searching for Atmospheric Species} \label{chap:searching}

Despite the removal of stellar and telluric lines using SysRem, the planetary signal remains faint and significantly buried below the photon noise level (as seen in the fourth panel of Figure \ref{fig:spec_anal}). Leveraging the benefits of high-resolution spectroscopy, which resolves numerous individual planetary lines, we enhance signal strength by combining these distinct lines through cross-correlation. This approach enables the robust detection of the planetary atmosphere, overcoming the challenges posed by its weak and buried nature within the data \citep{Birkby_2018}.


For every SysRem iteration, we cross-correlated each exposure of the data after telluric removal with the Doppler-shifted model spectrum at a velocity $v$ via
\begin{equation}\label{equ:CCF}
    \mathrm{CCF}(v) = \sum \frac{f_im_i(v)}{\sigma_i^2}
\end{equation}
where $f$ is the residual, $m$ is the model, and $\sigma$ is the residual uncertainty. Subscript $i$ denotes the $i$-th wavelength channel. The inclusion of $\sigma$ in the above equation ensures that proper weighting can be done since we might still have some uncorrected telluric residuals in the data. 

The spectrum model was Doppler-shifted in a grid of RV values from $-$350 to 350 $\mathrm{km\ s}^{-1}$ in a 1.3 $\mathrm{km\ s}^{-1}$ step (which is the average pixel step of CARMENES NIR spectrum) to account for the highly Doppler-shifted signal of the planet during the transit. For each spectral order, we cross-correlated each of the Doppler-shifted models with the data residuals and later summed up the calculated CCF over the orders. At this stage, the CCF matrix is at the telluric rest frame and has a dimension of the number of frames times the number of RV values in the grid.

To search for the planetary signal, we shifted the CCFs to the planetary rest frame using a linear interpolation via the circular orbit solution of the RV equation (since we expect that the orbit of hot gas giants had already been circularized), i.e.,
\begin{equation} \label{equ:RV_sin}
    RV_{\mathrm{p}}(t)=K_\mathrm{p}\sin{(2\pi \phi(t))}+V_{\mathrm{sys}}-V_{\mathrm{bary}}(t)+V_{\mathrm{rest}}
\end{equation}
where $K_\mathrm{p}$ is the semi-amplitude radial velocity of the planet, $V_{\mathrm{sys}}$ is the systemic velocity of the system, $V_{\mathrm{bary}}$ is the barycentric velocity during the observation (for CARMENES data, $V_{\mathrm{bary}}$ is given relative towards the observer, hence negative sign was used), $V_{\mathrm{rest}}$ is an additional term to account for any possible deviations from the expected $RV_\mathrm{p}$, and $\phi$ is the orbital phase which can be calculated via
\begin{equation}\label{phase}
    \phi=\frac{t-T_0}{P}
\end{equation}
where $t$ is time, $T_0$ is the transit midpoint or the inferior conjunction, and $P$ is the planetary orbital period. The equation is expressed such that $\phi=0$ corresponds to the transit midpoint. The theoretical $K_\mathrm{p}$ calculated using the planetary parameters in Table \ref{tab:stellar_planet_param} is $\sim164.27\pm0.83$ $\mathrm{km\ s}^{-1}$ (assuming zero eccentricity, which is expected for this planet; \citealp{Wolf_2007, Stevenson_2012}). The systemic velocity $V_{\mathrm{sys}}$ was measured by first cross-correlating the data with a PHOENIX model spectrum (with stellar parameters closest to HD 149026 given in Table \ref{tab:stellar_planet_param}), then we used the likelihood framework (see Section \ref{chap:likelihood}) to derive the best-fit stellar RV value for each exposure. We found a $V_{\mathrm{sys}}$ value of $-$$17.91\pm0.01$ km s$^{-1}$ which is consistent with literature within 1-$\sigma$ \citep{Kang_2011, Tabernero_2012, Gaia_2018, Ishizuka_2021} and thus was used as our expected $V_{\mathrm{sys}}$ value. 

We explored a grid of $K_\mathrm{p}$ and $V_{\mathrm{rest}}$ values from 0 to 300 $\mathrm{km\ s}^{-1}$ (in 2 $\mathrm{km\ s}^{-1}$ step) and $-$120 to 120 $\mathrm{km\ s}^{-1}$ (in 0.5 $\mathrm{km\ s}^{-1}$ step), respectively, and constructed the $K_\mathrm{p}-V_\mathrm{rest}$ map. For each value of $K_\mathrm{p}$ and $V_{\mathrm{rest}}$, we normalized the CCF of each exposure by subtracting its median value, then we summed the CCFs over the exposure weighted by the transit light-curve of the planet. The transit light curve was modeled using the physical and orbital parameters listed in Table \ref{tab:stellar_planet_param} with the publicly available \texttt{batman} package\footnote{\href{https://lkreidberg.github.io/batman/docs/html/index.html}{https://lkreidberg.github.io/batman/docs/html/index.html}} \citep{Kreidberg_2015}. We used the quadratic limb darkening model which coefficients calculated with ExoCTK Limb Darkening Calculator\footnote{\href{https://exoctk.stsci.edu/limb_darkening}{https://exoctk.stsci.edu/limb\_darkening}} using the stellar parameters also listed in Table \ref{tab:stellar_planet_param}. The calculated coefficients are 0.075 and 0.372 for $c_1$ and $c_2$ respectively.

We then estimated the S/N of the planetary signal following \cite{Brogi_2023} where we fit the noise distribution with a Gaussian profile and used the standard deviation of the fitted profile as the proxy to the true standard deviation of the map. The noise distribution excludes regions within $\pm150$ $\mathrm{km\ s}^{-1}$ in $K_\mathrm{p}$ and $\pm10$ $\mathrm{km\ s}^{-1}$ in $V_{\mathrm{rest}}$ from the strongest signal (i.e., the planet signal if it is present). We also excluded regions around the telluric rest frame ($0<K_\mathrm{p}<50$ $\mathrm{km\ s}^{-1}$) to avoid overestimating the resulting S/N as these regions are expected to have lower level of noise due to SysRem correction. Finally, we divided the $K_\mathrm{p}-V_{\mathrm{rest}}$ CCF map by the fitted standard deviation. This produced the S/N map (see next Section).

\section{Results} \label{chap:results}
\subsection{Evidence of water vapor} \label{chap:evidence} 
 
\begin{figure*}[t!]
\centering
\includegraphics[width=\textwidth,height=0.7\textheight,keepaspectratio]{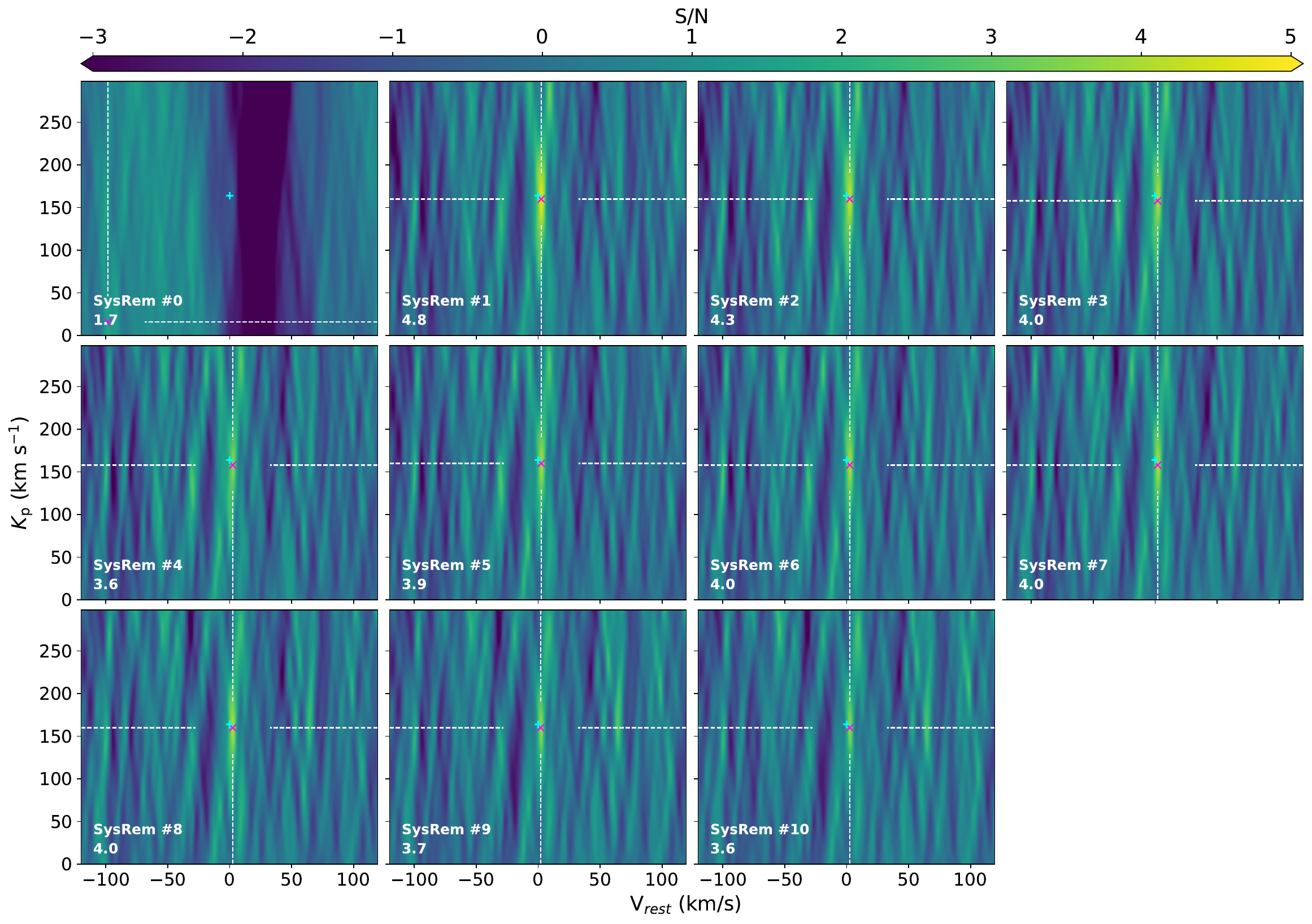}
\caption{S/N map for the POKAZATEL model. The $y$-axis is the semi-amplitude radial velocity while the $x$-axis is the systemic velocity measured in the planetary rest frame $V_\mathrm{rest}$. The white dashed lines point to the maximum peak in the map, marked by the red cross symbol, of which significance is indicated at the lower left corner of the map, along with the corresponding SysRem iteration The blue plus symbol marks the expected $K_\mathrm{p}$ and $V_\mathrm{rest}$ location of the planet.}
\label{fig:POKAZH2OCE_SN_map}
\end{figure*}

Figure \ref{fig:POKAZH2OCE_SN_map} is the S/N map after summing the CCF over time showing evidence of H$_2$O signal very close to the expected location of the planet which, interestingly, is red-shifted to $V_{\mathrm{rest}}\sim$ 2.5 $\mathrm{km\ s}^{-1}$. The strongest signal can be seen in the SysRem \#1 (i.e. first SysRem iteration) with an S/N of 4.8. As the number of SysRem iterations increases, the signal gets weaker (although we can see an unusual drop in SysRem \#4). If there was a significant contribution from telluric residuals, we would see a signal around the telluric rest frame at ($K_\mathrm{p}$, $V_{\mathrm{rest}}$) $\sim$ (0, $V_{\mathrm{bary}}$)\footnote{$V_{\mathrm{bary}}$ in this dataset ranges from 8.6 to 8.2 km s$^{-1}$ from start to end of observation.}. However, there are no signs of such signals both in the telluric rest frame and around the water signal, which adds robustness to the exoplanetary origin of the observed H$_2$O CCF peak. We also explored the unphysical negative $K_\mathrm{p}$ regions to search for potential false positives and found nothing that resembles the peak. Following the above results, all the remaining results to be presented in this paper are discussed in the context of the SysRem \#1. 

\begin{figure*}[t!]
\centering
\includegraphics[width=\textwidth,height=0.95\textheight,keepaspectratio]{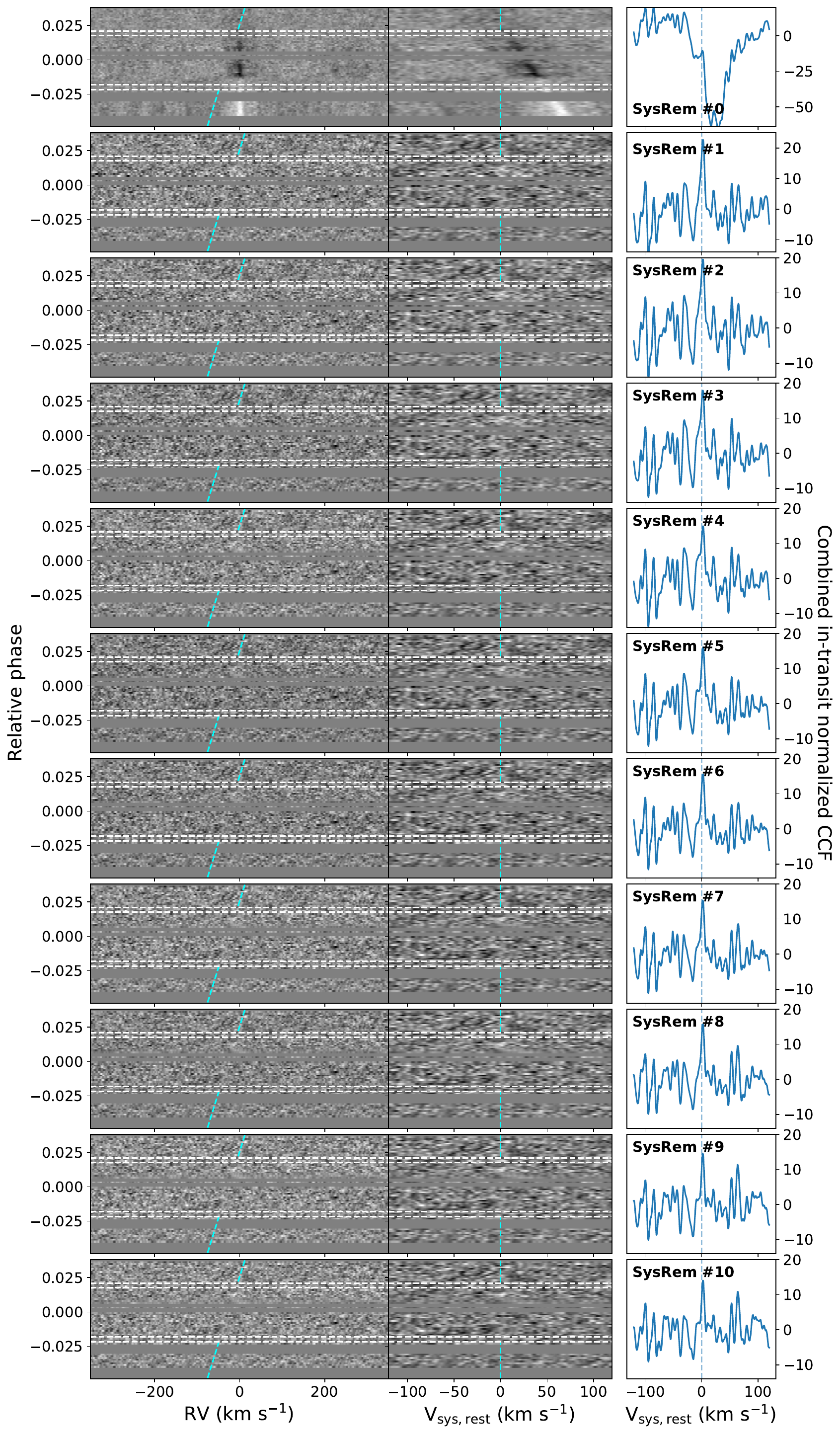}
\caption{CCF maps from cross-correlation between the unprocessed H$_{2}$O POKAZATEL model and SysRem residuals. Brighter color corresponds to higher values, i.e., a better match between the model and data. \textbf{Left}: CCF map as a function of RV values and the transit phase relative to the mid-point. The four horizontal dashed white lines are the estimated four transit contacts, starting from the bottom. The cyan dashed line locates the expected planetary RV path across the map. \textbf{Middle}: same as the left panel but the CCF has been shifted to the planetary rest frame and the horizontal axis is now the systemic velocity in planetary rest frame. \textbf{Right}: The combined in-transit CCF values from the middle panel. The blue vertical dashed line denotes the $V_\mathrm{sys, rest} = $ 0 $\mathrm{km\ s}^{-1}$ location. The zeroth SysRem iteration (before fitting for SysRem models) is also shown.}
\label{fig:POKAZH2OCE_CC_map}
\end{figure*}

For completeness, we show the CCF maps for the unprocessed POKAZATEL model for each SysRem iteration in Figure \ref{fig:POKAZH2OCE_CC_map}. We can see a peak close to the expected position at $V_{\mathrm{sys,rest}}\equiv V_{\mathrm{rest}}\approx0$ $\mathrm{km\ s}^{-1}$ in the combined CCF plot at the planet rest frame, showing our H$_2$O evidence. However, we also noticed a bright 'blob' around $\phi\sim0.0098$ which seems to even persist at SysRem \#10 and might contribute to a significant fraction of the CCF peak. We checked whether the CCF peak only comes from certain exposures or spectral orders (possibly indicating that it is just an analysis artifact) by removing the exposures and spectral orders one by one and then visually checking if the peak is still there. We confirmed that the CCF peak always persists with similar S/N in almost all cases except one, that is when we removed frame \#44 ($\phi\sim0.004$) which leads to a drop in S/N to 4.3, indicating that this frame might be a possible outlier source in the signal. Excluding this frame throughout our analysis does not significantly affect our results and conclusion, however, thus we decided to keep it for consistency. Putting it aside, the fact that we can still see the signal after subsequently removing the frame and spectral order one by one shows that the signal is the result of the combination of all exposures, not just from a single one.

In contrast, we also observe the signal when including all frames with S/N below 50, which were excluded in the primary analysis. Notably, the strongest signal in this case has an S/N of $\sim$4.2 and occurs at a higher SysRem iteration, specifically SysRem \#11 (with a total of 15 iterations conducted for this part). The lower S/N is primarily due to the signal peak having a lower CCF value before normalizing by the standard deviation, which tends to be higher for the same SysRem iteration. The increased number of required iterations is attributed to the inclusion of noisier data, namely exposures with S/N below 50, resulting in a slower removal of systematics by SysRem.

\begin{figure}[t!]
\centering
\includegraphics[width=\linewidth]{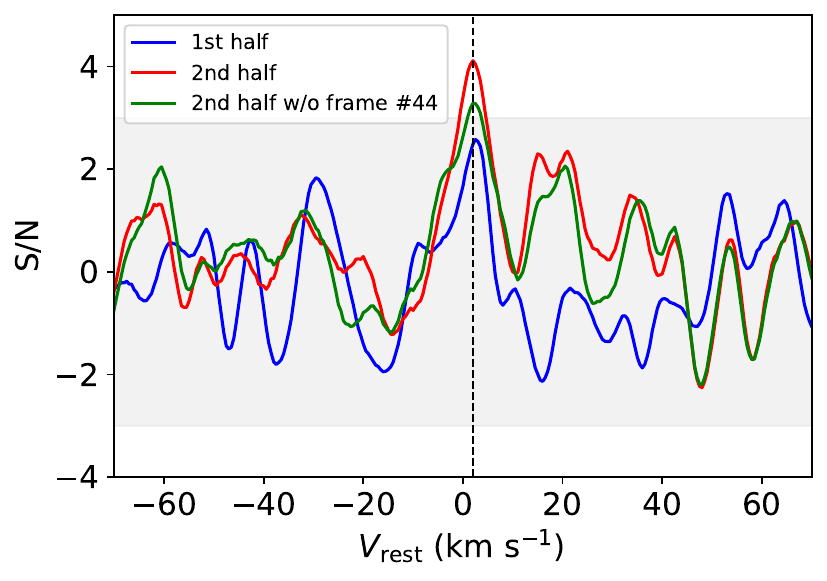}
\caption{Slices of S/N of H$_2$O POKAZATEL model at $K_\mathrm{p}=164.27$ $\mathrm{km\ s}^{-1}$ from the first (blue curve) and second half (red curve) of the transit. S/N for the second half without frame number \#44 is also shown (green curve). The vertical black dashed line corresponds to the retrieved $V_\mathrm{rest}$ from Figure \ref{fig:CEmodel_SNWelch_map} ($V_\mathrm{rest}=2.5$ $\mathrm{km\ s}^{-1}$). Gray area highlights regions within S/N $\pm$ 3.}
\label{fig:separate_half}
\end{figure}

At zeroth (\#0) SysRem iteration where telluric lines are still dominating, the telluric signal manifests itself as the dark and bright streak along the telluric rest frame. The wing of the telluric signal also seems to overlap with the expected planetary signal at the second half of the transit, especially the last phases, which could potentially introduce a spurious signal near the location of the planet. However, both transit halves seem to contribute to the detected signal, although in a different level like shown in Figure \ref{fig:separate_half} where we summed up cross-correlation for the first ($\phi<0$) and second half ($\phi>0$) separately. We found that the second half appears to show a stronger signal than the first half by $\sim$1.4 in S/N, as seen in Figure \ref{fig:separate_half}, showing that the second half contributes more to the detected signal. This might be partially explained by the poorer night condition during the first half both in airmass and relative humidity (Figure \ref{fig:night_condition}), obstructing more of the planet signal coming from this period. However, we note that, as has been discussed previously, frame \#44 ($\approx \phi\sim0.004$) seems to have quite a significant contribution to the signal. When we removed it from the second half, we got a weaker signal down to roughly 3.3 in S/N (green curve in Figure \ref{fig:separate_half}) comparable to the first half whilst the combined S/N becomes 4.3, 0.5 lower than the original signal we detected. In addition, telluric residuals are expected to be minimal after applying even one SysRem iteration (Figure \ref{fig:telluric_ccf}). Thus, we might expect that the overlapping telluric wing does not produce any signal at the first SysRem iteration onwards.

\begin{figure*}
\centering
\includegraphics[width=0.9\linewidth]{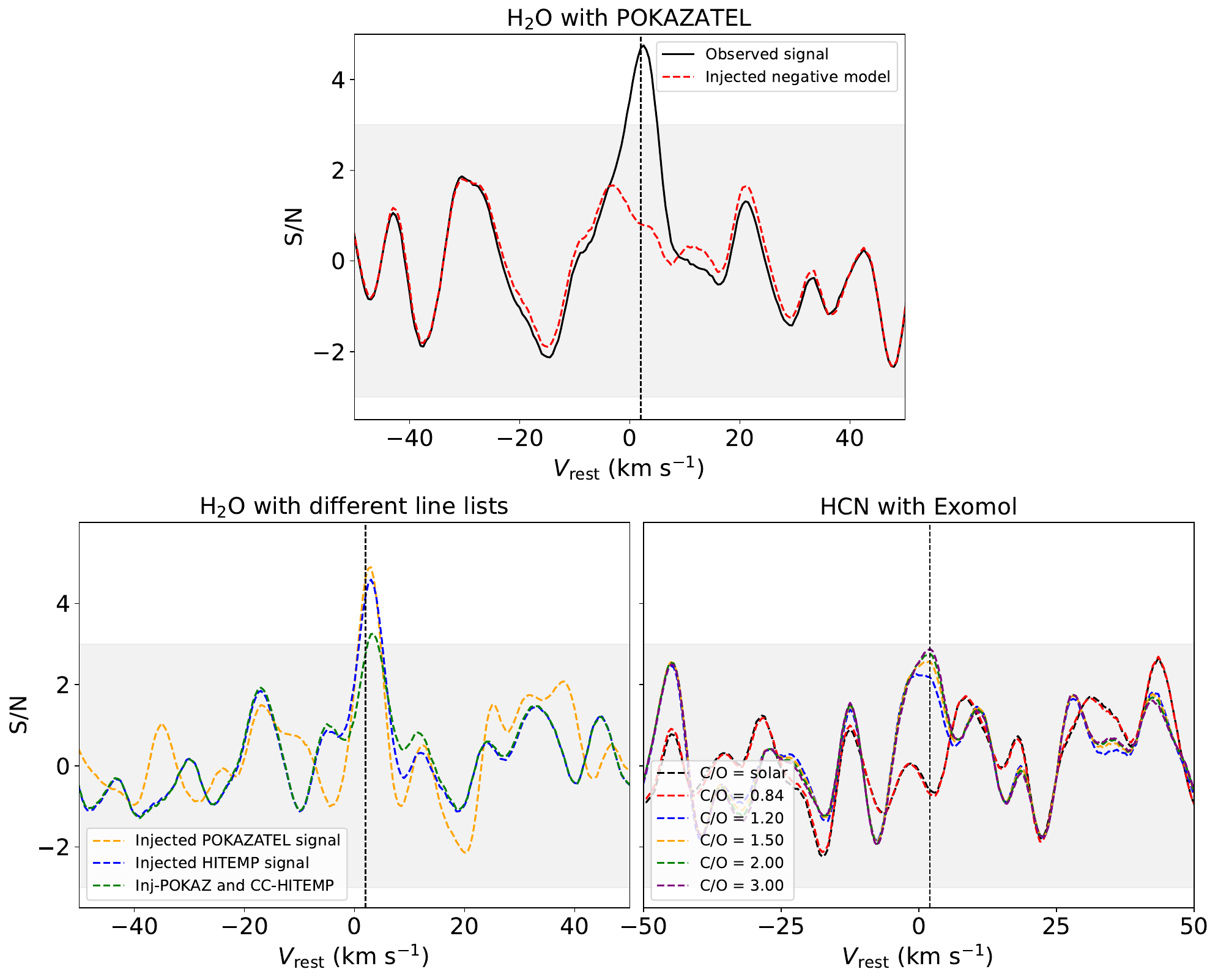}
\caption{Slices of the S/N map at $K_\mathrm{p}=164.27$ $\mathrm{km\ s}^{-1}$ for upper panel and $K_\mathrm{p}=-164.27$ $\mathrm{km\ s}^{-1}$ for lower panels for various injection tests with injected models shown by the legend. All injected water (POKAZATEL and HITEMP) models are injected at the detected $V_\mathrm{rest}$. Regions within S/N $\pm$ 3 are marked by the gray color. Black vertical dashed lines mark the $V_\mathrm{rest}$ used for models in each panel.}
\label{fig:inject_CEmodel}
\end{figure*}

We made a simple test to assess whether our POKAZATEL signal is real or just an analysis artifact. We did this by injecting a negative POKAZATEL model at the detected $K_\mathrm{p}$ and $V_{\mathrm{sys}}$ just before putting the continuum to a common level (we assumed that, given a sufficiently wide window and standard deviation, the continuum correction procedure will not significantly alter the planet signal). We injected the model at 0.96 times the nominal strength (this is based on the retrieved $\alpha$ in Figure \ref{fig:3DMCMC_equ}; see Section \ref{chap:likelihood}). We then performed the exact same analysis as has been described in previous sections. The red dashed curve in the upper panel of Figure \ref{fig:inject_CEmodel} shows the slice of the resulting S/N map at $K_\mathrm{p}=164.27$ $\mathrm{km\ s}^{-1}$. For comparison, the black solid curve is the observed H$_2$O signal from POKAZATEL. If the detected signal is not real (i.e. noise or only coming from several exposures or spectral order) we would have seen a negative peak. Instead, the H$_2$O signal seems to disappear completely once we inject the negative model, resulting in a close-to-zero S/N, strengthening the evidence of its planetary origin.

\begin{figure*}
\centering
\includegraphics[width=0.8\linewidth]{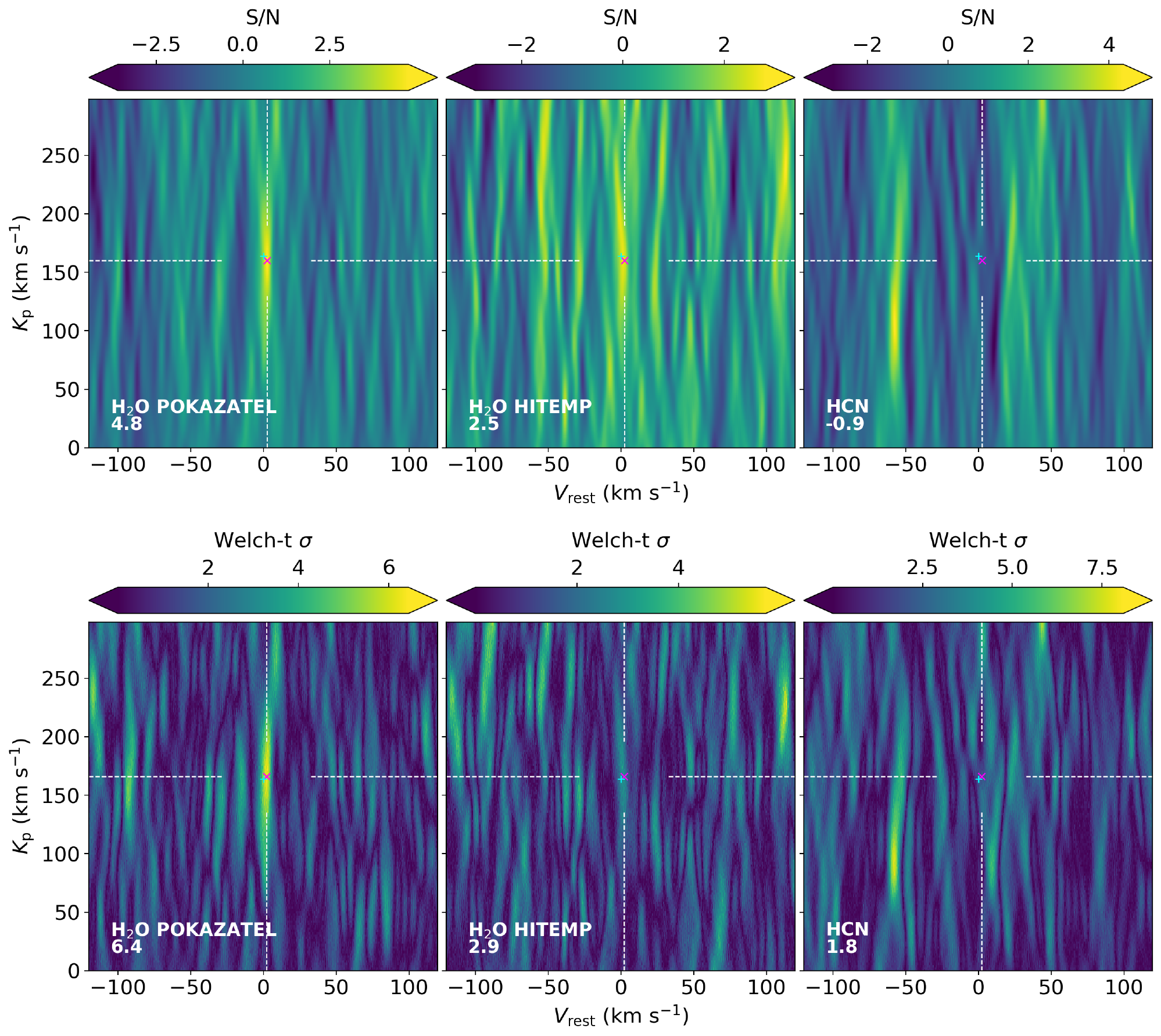}
\caption{\textbf{Top}: S/N map for each of the unprocessed models. The white dashed lines point to the detected $K_\mathrm{p}$ and $V_\mathrm{rest}$ values from the H$_2$O POKAZATEL model, ($K_\mathrm{p},V_\mathrm{rest}$) $\sim$ (160.0, 2.5) $\mathrm{km\ s}^{-1}$, marked by the red cross symbol with significances indicated at the lower left corner of each map. These significance values are evaluated at the location of the H$_2$O POKAZATEL signal. The blue plus symbol marks the expected planet location. \textbf{Bottom}: Same as the top panel but for Welch-t $\sigma$ with in-trail width of 3.73 $\mathrm{km\ s}^{-1}$ (FWHM of CARMENES). All maps are derived from SysRem \#1. We note that we cannot assess the optimum SysRem iteration for the HCN model since this species cannot be detected.}
\label{fig:CEmodel_SNWelch_map}
\end{figure*}

The top panels of Figure \ref{fig:CEmodel_SNWelch_map} summarize our results from the S/N maps of the unprocessed models. Whilst we found evidence of water from the POKAZATEL line list, surprisingly we could not find the same signal using the HITEMP line list at least at a similar S/N. We also could not detect HCN. The non-detection of H$_2$O HITEMP and HCN will be discussed further in Section \ref{chap:line list} and \ref{chap:nonHCN}, respectively.

\subsection{Welch-t test} \label{chap:welch-t}
To further assess the robustness of the detection, we also performed the Welch-t test, which is commonly used in literature \citep[e.g.][]{Nugroho_2017, Sanchez-Lopez_2019, Webb_2020, Giacobbe_2021}. The test is used to determine whether two distributions are coming from the same parent distribution by evaluating whether they have the same mean value. In this case, these two distributions are the planet signal and the noise called the in-trail and out-trail, respectively. The null hypothesis of this test is that these two distributions have the same mean and the corresponding \textit{p-value} will measure the significance at which the null hypothesis is rejected, i.e., how far the mean of the in-trail distribution is shifted from the out-trail distribution. One important note is that this test is only valid if the two distributions are normally distributed. We found that, in our analysis, the out-trail and in-trail distribution are Gaussian to roughly 4-$\sigma$ and 2.5-$\sigma$, respectively. 

\begin{figure}
\centering
\includegraphics[width=\linewidth]{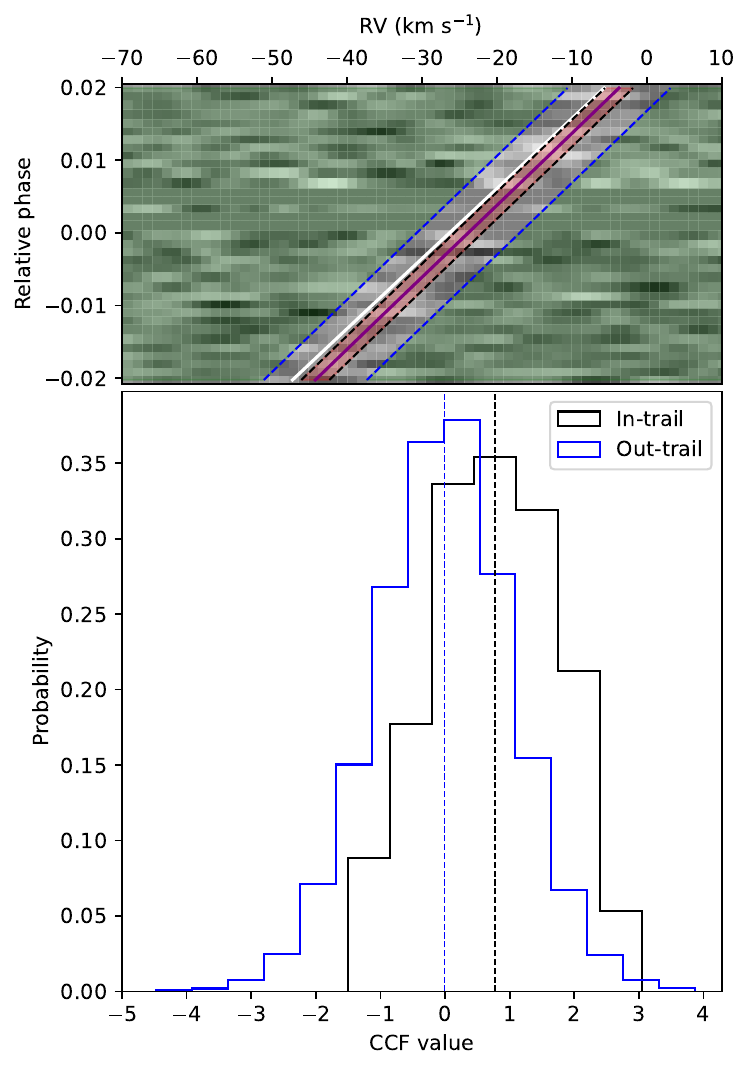}
\caption{\textbf{Top}: the in-transit section of left panel of Figure \ref{fig:POKAZH2OCE_CC_map} for SysRem \#1. The white solid line is the expected planetary RV path whilst the purple solid line is the planetary RV path corresponding to the detected $K_\mathrm{p}$ and $V_\mathrm{rest}$ values from Figure \ref{fig:CEmodel_SNWelch_map}. The red area is the in-trail distribution and spans $\pm$ $\sim$3 pixels (with 1.3 $\mathrm{km\ s}^{-1}$ sampling) from the expected planetary path (red solid line). The green area is the out-trail distribution. The area between the in-trail and out-trail that spans 5 $\mathrm{km\ s}^{-1}$ (from the black dashed lines to the blue dashed lines) is excluded from the analysis. \textbf{Bottom}: the histogram for both the in-trail and out-trail distribution. The vertical dashed lines are the mean of each distribution, calculated by fitting a Gaussian to the histogram and subsequently computing its mean, with colors shown by the legend.}
\label{fig:welch_dist}
\end{figure}

The bottom panel of Figure \ref{fig:welch_dist} shows the histogram of both distributions drawn from the top panel of the Figure (which is the in-transit CCF map from the left panel of Figure \ref{fig:POKAZH2OCE_CC_map} for SysRem \#1 ). The in-trail distribution includes the CCF values inside the black dashed lines while the out-trail includes all of the CCF values outside the blue dashed lines (lines separating the out-trail regions from the rest of the map). We can see that the mean value of the in-trail distribution is shifted from 0, which is the mean for the out-trail. We then calculated the \textit{p}-value and converted half of the value to detection significance using inverse survival function\footnote{\href{https://docs.scipy.org/doc/scipy/reference/generated/scipy.stats.norm.html}{scipy.stats.norm.isf}}, following \cite{Giacobbe_2021}, assuming unequal variances. We performed this procedure for each combination of $K_\mathrm{p}$ and $V_{\mathrm{rest}}$ and each chemical species searched in this work. The resulting Welch-t maps are shown by the lower panels of Figure \ref{fig:CEmodel_SNWelch_map} assuming an in-trail width of 3.73 $\mathrm{km\ s}^{-1}$. 

\begin{figure}
\centering
\includegraphics[width=\linewidth]{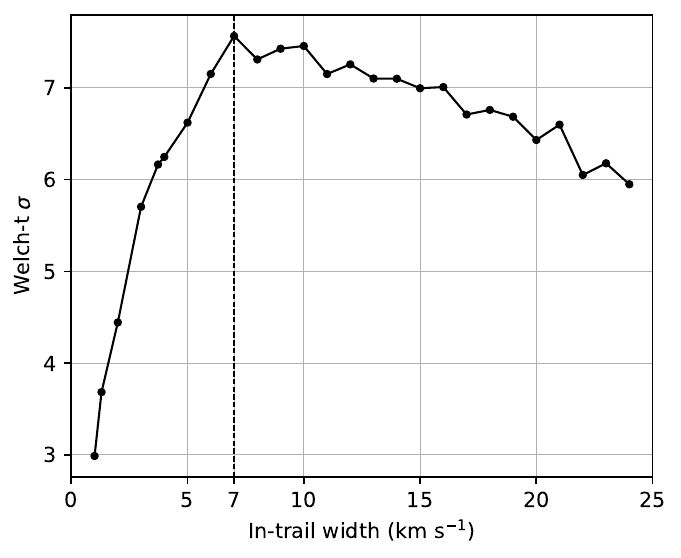}
\caption{Welch-t $\sigma$ evolution with the in-trail width. The vertical dashed line marks the location of maximum significance. Significance is evaluated at a fixed value of $K_\mathrm{p}$ and $V_\mathrm{rest}$ taken from the detected signal location in the S/N map of POKAZATEL in Figure \ref{fig:CEmodel_SNWelch_map}.}
\label{fig:intrail_width}
\end{figure}

The strongest signal from the POKAZATEL model (with an in-trail width of 3.73 $\mathrm{km\ s}^{-1}$) was found at $K_\mathrm{p}\sim$ 166.0 $\mathrm{km\ s}^{-1}$ and $V_{\mathrm{rest}}\sim$2.0 $\mathrm{km\ s}^{-1}$, close to the expected planet location, at a significance of 6.4-$\sigma$. Similar to S/N map results, no significant peak appeared for the H$_2$O HITEMP model and HCN around the expected location. Although the POKAZATEL signal location is consistent with that retrieved from the S/N map, the significance itself is higher by roughly 1.6-$\sigma$. However, we do note here that the retrieved significance highly depends on the in-trail width. As the in-trail width increases, more planet signals are included in the in-trail distribution, hence boosting the detection significance. That is until we get to a point where the significance reaches a maximum and instead decreases afterward due to more noise included in the in-trail rather than the planet signal itself \citep{Webb_2020}. This is illustrated in Figure \ref{fig:intrail_width} which shows how the significance reaches a maximum when the in-trail width is 7 $\mathrm{km\ s}^{-1}$ then gets lower and is expected to keep gradually decreasing as the in-trail width increases. Hence, if we were to adopt the maximum significance as our in-trail width, we would have a much larger velocity broadening than expected, as we will discuss next.

Since in-trail distribution should include the planet signal, we would expect that the highest significance is reached once the in-trail width roughly corresponds to the width of the CCF profile of the signal. This then suggests the detected signal has a wide line broadening ($\sim$7 $\mathrm{km\ s}^{-1}$) which significantly exceeds the FWHM of the data ($\sim$3.73 $\mathrm{km\ s}^{-1}$) and the expected rotational broadening of the planet ($\sim$1.43 $\mathrm{km\ s}^{-1}$). However, we cannot conclude anything about this solely from the Welch-t test  as the in-trail width does not directly relate to the line FWHM. Further atmospheric retrieval to constrain the broadening velocity is needed in future work.

Due to this high dependency and the rather arbitrary in-trail width choice, Welch-t test produces less reliable results in our analysis and we encourage the reader to refer more to the S/N maps instead.

\subsection{Comparing different H$_{2}$O line lists} \label{chap:line list}
The discrepancy on S/N between the POKAZATEL and HITEMP water line list is, at first, expected as similar results were also found in the literature \citep[e.g.][]{Gandhi_2020, Webb_2020, Nugroho_2021}. \cite{Nugroho_2021}, for example, managed to marginally detect H$_2$O using POKAZATEL and HITEMP (although much weaker for the latter), but could not find anything using BT2 \citep{Barber_2006}. This may be caused by differences in the line strength and position between line lists, resulting in different models, even with the same atmospheric parameters. A prime example of this can be seen from the water models at around 13,341 \AA, where we can see a very strong absorption line in the HITEMP model (peaked to roughly 0.94 $R_\mathrm{Jup}$), but not in the POKAZATEL model. However, the fact that our results show a significant difference in S/N between H$_2$O from HITEMP and POKAZATEL is surprising.

To see whether the difference between line lists is indeed expected, we injected both the POKAZATEL and HITEMP models in the same manner as has been described in Section \ref{chap:evidence}, however this time the models are injected at the negative value of the expected $K_\mathrm{p}$ (to avoid contamination from the real planet signal; \citealp{Gibson_2020, Gibson_2022, Merritt_2020}) and scaled to 1.4 times the nominal strength (this choice is semi-arbitrary and was chosen because it resulted in a similar S/N with the observed POKAZATEL signal). The orange and blue dashed lines in the lower left panel of Figure \ref{fig:inject_CEmodel} shows the result for the injected POKAZATEL and HITEMP model respectively. As we can see, there is a slight discrepancy between the two line lists, with a maximum HITEMP signal of S/N $\sim$4.5 and a maximum POKAZATEL signal of $\sim$4.9. Although we see higher significance for the injected POKAZATEL model, the difference is slight and the opposite of what we see from the observed signal where the POKAZATEL signal shows almost double than the HITEMP signal. We note, however, that the value of the S/N of the recovered H$_2$O signal will depend on the injected location. For example, if we instead injected the models at $V_{\mathrm{rest}}=0$ $\mathrm{km\ s}^{-1}$, rather than $V_{\mathrm{rest}}=2.5$ $\mathrm{km\ s}^{-1}$, we would see that the recovered HITEMP signal would be slightly stronger than POKAZATEL. In addition, injecting models at different locations would also result in different optimal SysRem iterations from that of the real data. For example, for our injected POKAZATEL and HITEMP model here, the recovered injected signal reaches a maximum after the third and sixth iteration, respectively. All of these are because different locations in the $K_\mathrm{p}-V_{\mathrm{rest}}$ map would have different noise behavior, leading to different CCF (or S/N) values (this is also why we need a stronger injected model at the negative $K_\mathrm{p}$ to match the observed S/N of the POKAZATEL signal). Hence, the S/N recovered from the injection test should not be interpreted at face value, and choosing the optimum SysRem iteration using the injection test alone is not recommended.

We also tried to recover injected POKAZATEL model (at negative $K_\mathrm{p}$) using the HITEMP model to qualitatively examine how different the two models are. The result is shown by the green dashed curve in the lower left panel of Figure \ref{fig:inject_CEmodel} where there is a peak around the expected $V_{\mathrm{rest}}\sim2.5$ $\mathrm{km\ s}^{-1}$ at an S/N of $\sim$3.1, $\sim$0.6 higher in S/N to that with the observed HITEMP signal. This might indicate that the underlying water signal in the data is best represented by the POKAZATEL model rather than HITEMP itself given the adopted atmospheric parameters. Note that, other than the line list itself, we used identical atmospheric parameters for both models.

\subsection{Non-detection of HCN} \label{chap:nonHCN}
Contrary to H$_2$O POKAZATEL, we could not detect HCN in any SysRem iterations, though we see some peaks in the HCN map around $V_{\mathrm{rest}}= -50$ $\mathrm{km\ s}^{-1}$ spanning over a wide range of $K_\mathrm{p}$. However, these are likely only systematic noises since they are too far from the expected planet location. We performed injection tests for HCN with different C/O ratios to see whether we could detect HCN if it had a sufficiently high abundance. We injected five HCN models with C/O ratios ranging from solar (C/O = 0.55) to 3.00 in the same manner as with the injected water models at negative $K_\mathrm{p}$ (see previous section) except that we injected them at nominal strength (scaled by 1). The lower right panel of Figure \ref{fig:inject_CEmodel} shows the results. 

For C/O $<$ 1 where the abundance of HCN is     low, we cannot recover any signal at the planet location even for the expected C/O ratio of this planet (C/O = 0.84; \citealp{Bean_2023}). As the C/O increases from 0.84 to 1.20, there is an extreme jump of $\sim2.9$ in S/N from a negative value to $\sim$ 2.1. This is where the abundance of HCN dominates, surpassing the H$_2$O abundance, and the atmosphere becomes carbon-dominated \citep{Molliere_2015}. Although the S/N of the peak keeps increasing as we increase the C/O ratio, the recovered signal is still weak even for a C/O ratio as high as 3.00 (S/N = 2.9; not high enough to typically claim a detection). We strongly note that the recovered S/N will highly depend on the injected model strength, which in this case is the nominal strength. However, the fact that nominal model strength for even C/O of 3.00 does not result in a favorable S/N suggests that our data are not sensitive to HCN even if it was present in the atmosphere of the planet.

\pagebreak
\subsection{Orbital parameter grid search} \label{chap:likelihood}

\cite{Brogi_2019} showed that a Bayesian approach can be applied for high-resolution analysis to retrieve the planetary parameters. \cite{Gibson_2020} then developed the generalized version of \cite{Brogi_2019} likelihood by incorporating the uncertainty of each data point and a scaling factor $\beta$ for the uncertainty. Here, we specifically used the $\beta$-optimized version of the likelihood which is expressed as
\begin{equation}\label{equ:logL}
\begin{aligned}
\ln{(\mathcal{L})} = & -\frac{N}{2}\ln\left[\frac{1}{N}\left(\sum\frac{f_i^2}{\sigma_i^2} + \right. \right. \\
      & \left.\left. \alpha^2\sum\frac{m_i^2}{\sigma_i^2}-2\alpha\sum\frac{f_i m_i}{\sigma_i^2}\right)\right]
\end{aligned}
\end{equation}
where $N$ is the number of data points and $\alpha$ is a nuisance parameter that essentially acts as the scale factor for the model. The advantage of using $\alpha$ is that it will correct the model scaling in the case where the model is either underestimating or overestimating the observed signal, especially when we use a simple model. Note that the third term inside the bracket is the CCF shown by Equation \ref{equ:CCF}.

\begin{figure}[t!]
\centering
\includegraphics[width=\linewidth]{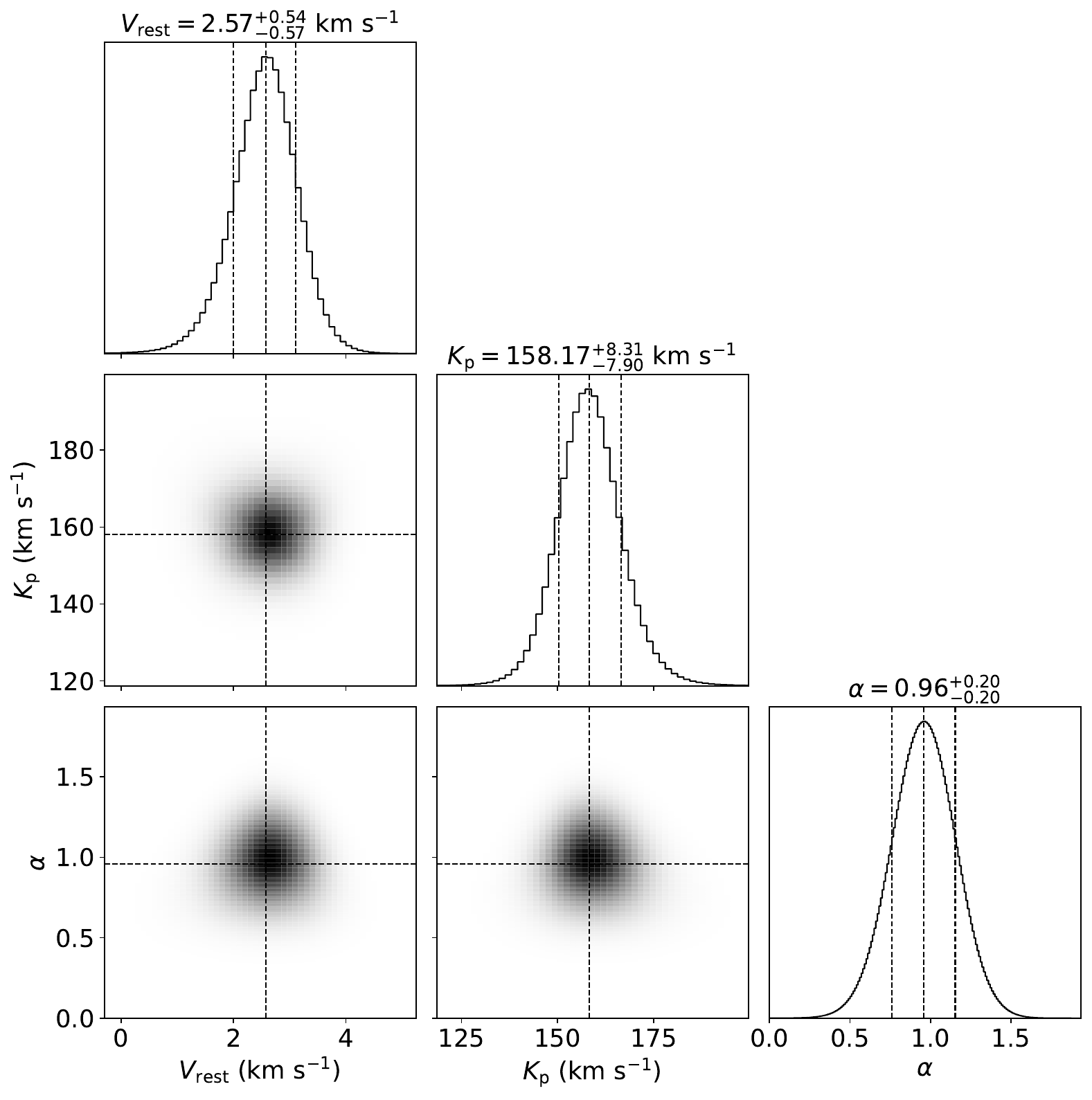}
\caption{Corner plot of the normalised likelihood $\mathcal{L}$ values for the fit parameters. Vertical dashed lines in the marginalised distribution panels correspond to the 16-th, 50-th, and 84-th percentiles from left to right.}
\label{fig:3DMCMC_equ}
\end{figure}

We performed a grid search to explore the likelihood for $K_\mathrm{p}$, $V_{\mathrm{rest}}$, and $\alpha$ through the following procedure. For each parameter combination, we first pre-processed the POKAZATEL model, then Doppler-shifted the model according to the orbital phases. We then cross-correlate the model with the residuals and map the corresponding CCF value to log-likelihood via Equation \ref{equ:logL}. We explored a grid range of 50 to 250 $\mathrm{km\ s}^{-1}$ (with a step of 1.6 $\mathrm{km\ s}^{-1}$), $-$10 to 10 $\mathrm{km\ s}^{-1}$ (with a step of 0.1 $\mathrm{km\ s}^{-1}$), and 0 to 3 (with a step of 0.01) for $K_\mathrm{p}$, $V_{\mathrm{rest}}$, and $\alpha$, respectively. Figure \ref{fig:3DMCMC_equ} shows the result. We constrained the $K_\mathrm{p}$ to $158.17^{+8.31}_{-7.90}$ $\mathrm{km\ s}^{-1}$, consistent with the expected value within 1-$\sigma$ and is overlapping with the error bar of the retrieved $K_\mathrm{p}$ for neutral titanium signal from \cite{Ishizuka_2021}. Meanwhile, the retrieved $V_{\mathrm{rest}}$ is highly red-shifted to 2.57$^{+0.54}_{-0.57}$ $\mathrm{km\ s}^{-1}$ at more than 3-$\sigma$ away from the expected value. Possible explanations and consequences of this red-shifted velocity will be discussed further in Section \ref{chap:discussion}. 

The retrieved $\alpha$ shows that the model slightly overestimates the data and needs to be scaled up to 0.96$^{+0.20}_{-0.20}$. This might tell us that our chemical equilibrium model might not be the actual representative of the planetary atmosphere. We note that $\alpha$ only acts as a scaling factor for the model, which means that all spectral lines in the model are scaled up/down the same way. Since model parameters, specially the temperature, affect each spectral line differently, thus $\alpha$ cannot, in any way, mimic the effect given by different model parameters. 

Based on $\alpha$ definition, where a perfect match between data and model results in $\alpha = 1$, and an absence of planetary signal yields $\alpha = 0$ (indicating no detection), it is possible to estimate the detection significance using $\alpha$ \citep{Gibson_2020, Nugroho_2020a, Nugroho_2020b, Nugroho_2021}. This estimation is performed by dividing the median $\alpha$ value from the conditional likelihood (taken at the location of the maximum signal) by its standard deviation. We determined the $\alpha$-significance to be approximately 4.8-$\sigma$, which aligns with the detected S/N from Figure \ref{fig:CEmodel_SNWelch_map} (although the two metrics are not directly comparable due to different analysis treatments). Notably, detection significance calculated from $\alpha$ provides a more robust result. Unlike the S/N map, where detection significance depends on the range of $K_\mathrm{p}$ and $V_{\mathrm{rest}}$ used (as the calculated standard deviation varies with grid range), or the Welch-t significance, which is highly sensitive to in-trail width, this method is minimally biased by analysis parameters. It might also be noteworthy that the $\alpha$-significance decreases by 0.5-$\sigma$ in the absence of pre-processing.

\begin{figure*}
\centering
\includegraphics[width=\linewidth]{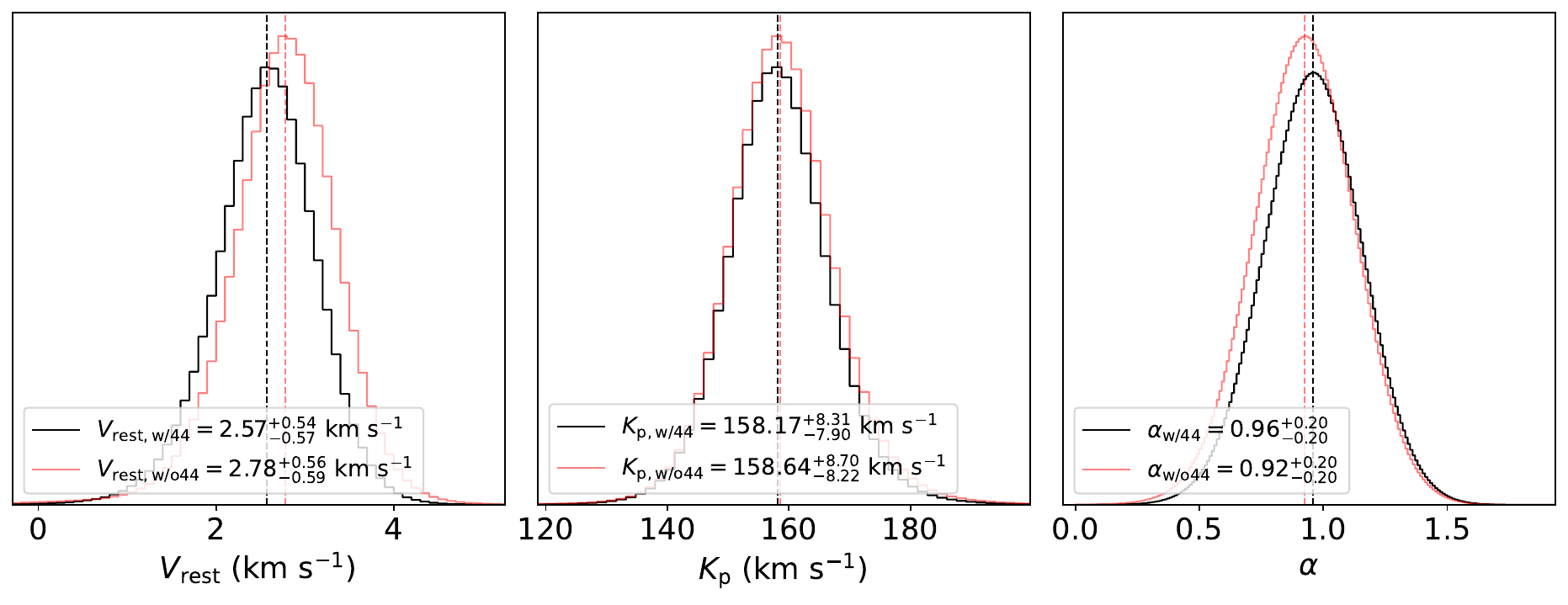}
\caption{Marginalised posteriors for $V_\mathrm{rest}$ (left), $K_\mathrm{p}$ (middle), and $\alpha$ (right) using the H$_2$O POKAZATEL model. The vertical dashed lines mark the median values that are indicated in the legend alongside the $\pm$1-$\sigma$ standard deviation. For each parameter, three posteriors are given where black indicates the case where we include frame \#44 and red indicates the case where we exclude it.}
\label{fig:posterior_w44_wo44}
\end{figure*}

We also check our seemingly biased results in cases where we are both including and excluding frame \#44 in context of the grid search. This is to see whether we would see a significant difference in the retrieval result between the two cases since we have seen in Figure \ref{fig:separate_half} that this particular frame might be quite a strong outlier. We conducted the same grid search setup using the H$_2$O POKAZATEL model. Figure \ref{fig:posterior_w44_wo44} shows marginalised posteriors for the three free parameters for the case where we include (black) and exclude (red) frame \#44. It can be seen that the results are robust against this frame with differences between the posteriors for all parameters are all within 1-$\sigma$. Note that the slight change in $\alpha$ from 0.96 to 0.92 resulting in a $\sim$0.2-$\sigma$ difference in the $\alpha$-significance between both cases. This difference is smaller than that from the S/N map ($\sim$0.5; see Section \ref{chap:evidence}), showing the robustness in using $\alpha$-significance compared to that of S/N.

\section{Discussion} \label{chap:discussion}
\subsection{On the C/O ratio and metallicity}
The evidence of H$_2$O suggests that the planet C/O ratio must be less than one if the atmosphere is homogeneous and at chemical equilibrium as the HCN abundance is expected to be very low \citep{Molliere_2015}. This is consistent with \cite{Bean_2023} which constrained the C/O of the day-side atmosphere to 0.84 $\pm$ 0.03 (although our result cannot rule out solar C/O ratio) but contradicts \cite{Ishizuka_2021} which suggests a C/O $\gtrsim$ 1 due to their non-detection of TiO. 

One important thing to note is that we are probing the terminator region of the planet unlike \cite{Bean_2023}. Since the composition of the day- and night-side of such hot gas giant might be significantly different, indicated by the planet's inefficient day-to-night heat redistribution and a temperature contrast that reaches $\sim$800 K \citep{Zhang_2018,Gagnebin_2024}, this means that we might not get the same C/O ratio through transmission observations as that found by \cite{Bean_2023}. We also strongly note that our data might not be sensitive enough to detect HCN even with a C/O ratio as high as 3.00 (see Section \ref{chap:nonHCN}). It is possible, however, to detect both H$_2$O and HCN, such as shown by \cite{Giacobbe_2021} and \cite{Sanchez-Lopez_2022}. This is because, in an oxygen-dominated atmosphere, quasi-continuum of H$_2$O would likely to hinder the much weaker HCN signal and hence, it is possible to detect both molecules only if HCN and H$_2$O stem from different limbs of the planet. Indeed, more data are needed to confirm our non-detection of HCN.

Recent study from \cite{Gagnebin_2024} analyzed the same JWST/NIRCam emission dataset with that of \cite{Bean_2023}. They used grid-retrieval technique (instead of free retrieval as in \citealp{Bean_2023}) and constrained the metallicity to [M/H]$=1.16^{+0.26}_{-0.37}$ if VO does not present in the atmosphere and [M/H]$=1.31^{+0.18}_{-0.23}$ if it does. Either way, both results are significantly lower than the constrained values from \cite{Bean_2023} with [M/H]$=2.09^{+0.35}_{-0.32}$, likely to be caused by the difference of the (retrieved) T-P profile shape between the two studies \citep{Gagnebin_2024}. Following this, we perform a simple analysis to find the preferred model metallicity for our data by performing the grid search like discussed in Section \ref{chap:likelihood} for a range of $-4\le\mathrm{[M/H]}\le4$ with a step of 0.1 in dex and fix the isothermal temperature to 1700 K as well as $K_\mathrm{p}$, $V_\mathrm{rest}$, and $\alpha$ to the retrieved values as shown in Figure \ref{fig:3DMCMC_equ}. We found that our data is not particularly sensitive to the metallicity as we can only loosely constrain it to [M/H]$=0.32^{+0.95}_{-0.80}$, making it to be consistent with both the host star's metallicity ([M/H]$=0.36$) and the unlikely sub-solar metallicity values \citep{Fortney_2006,Fortney_2013}. However, it should be noted that this result should not be taken at face value since we are not performing a full retrieval (taking the velocities, temperature, and cloud pressure as free parameters) as it is outside the scope of this work, hence further follow-up transmission observations are needed to confirm the water signal that we found and subsequently constrain the metallicity.


\subsection{Red-shifted H$_2$O} \label{chap:redshifted_vsys}

Whilst Doppler-shifted signals relative to the systemic velocity are commonly found \citep[e.g.][]{Snellen_2010, Sanchez-Lopez_2019, Nugroho_2020a, Yan_2023}, such shifts are usually caused by the very strong day-to-night side winds of tidally-locked hot gas-giants which typically is in order of a few $\mathrm{km\ s}^{-1}$. During transit, these winds are moving to the observer line-of-sight and combined with the planetary spin it will cause an overall blue-shift in the signal with magnitudes getting stronger as the transit progresses \citep{Showman_2013, Wardenier_2021} that also manifest as a shifted $K_\mathrm{p}$ to a smaller value than the expected \citep[e.g. ][]{Prinoth_2022}. For reference, \cite{Ishizuka_2021} found that the neutral titanium signal is blue-shifted to $-3.2\pm0.4$ $\mathrm{km\ s}^{-1}$. Although the titanium signal might probe different layers than our signal here (hence different atmospheric dynamics), our red-shifted signal is a bit anomalous. The shift in the water signal is unlikely to be caused by telluric lines as we have shown in Figure \ref{fig:telluric_ccf} that these lines are well removed even at the first SysRem iteration. In addition to cross-correlating SysRem residuals with telluric model and summing the CCF up over telluric rest frame (Section \ref{chap:SYSREM}), we also did the same but this time the cross-correlation was done with the planet model to see whether telluric residuals could be produced from cross-correlating the planet model. We could as well confirm that no significant telluric residuals were seen after performing one SysRem iteration. We also constrained the systemic velocity of our data of which value is consistent with literature \citep{Kang_2011, Tabernero_2012, Gaia_2018, Ishizuka_2021} and used it throughout the analysis described here (see Section \ref{chap:searching}). It would need a significant $V_\mathrm{sys}$ deviation ($>5\sigma$) to cause a 2.57 $\mathrm{km\ s}^{-1}$ red shift in $V_\mathrm{rest}$, hence it is unlikely that the adopted systemic velocity causes the derived $V_{\mathrm{rest}}$ to be red-shifted.

Another possible explanation is that the water signal is stronger in, or solely originates from, the morning terminator of the planet where winds flowing from the night- to day-side could exist, similar to what is observed for HCN in the atmosphere of WASP-76 b \citep{Sanchez-Lopez_2022}. If this is true, we would see a stronger signal during the first half, especially around the ingress, since the morning terminator should have a larger contribution to the signal during these phases. As the transit progresses, the evening terminator would begin to contribute more to the overall signal, weakening the overall water signal coming possibly mostly from the morning terminator. However, whether we could see this limb variations or not would depend on the angle subtended by the planet during transit and the opening angle; the extent of the atmosphere that contributes to the observed transmission spectrum \citep{Wardenier_2021}. If the latter is smaller than the former, as in the case for ultra-hot Jupiters such as WASP-76 b \citep[e.g.][]{Sanchez-Lopez_2022} due to their extreme proximity to their host stars, then the variations between the start and end of the transit would be more prominent and hence is easier to see. Following Equation 10 in \cite{Wardenier_2021}, in the case for HD 149026 b, the opening angle ranges from roughly 24 to 13 degrees for $0.01 \leq \beta \leq 0.1$ (fraction of absorption at a certain location to the total absorption along the transit chord by the species of interest). Since the planet sweeps $\approx$ 17 degrees of its orbit during transit, hence we might or might not be able to see the limb variations. Especially for our dataset, it would be more difficult because of its relatively low S/N and that we do not have sufficient amount of exposures during ingress and egress where we would expect the variations to be the greatest.

A similar overall red-shifted signal has been observed on the day-side of the hot Jupiter $\mathrm{\tau}$ Boötis b \citep{Pelletier_2021, Webb_2022}. \cite{Webb_2022} demonstrated that this red shift can be partially explained by adopting an eccentric rather than a circular orbital solution (see also \citealp{Montalto_2011}). For HD 149026 b, although the orbit is likely circular \citep{Wolf_2007, Stevenson_2012}, several studies \citep{Knutson_2009,Winn_2008,Knutson_2014,Bonomo_2017,Ment_2018} have found a non-zero but near-zero eccentricity, with some only providing the lower limit $e\cos{\omega}$. This might account for our observed red-shifted velocity, as a circular solution was used throughout this work. For instance, \cite{Ment_2018} constrained the eccentricity to $e = 0.051 \pm 0.019$, which could result in approximately a 2.5 $\mathrm{km\ s}^{-1}$ red shift in $V_{\mathrm{rest}}$, assuming the same argument of periastron $\omega$ for both eccentric ($e = 0.051$) and circular ($e = 0$) orbits. In light of the more recent JWST data, \cite{Bean_2023} provides two secondary eclipse times with about one-minute precision (although a circular orbit is assumed). Extrapolating the previously constrained transit midpoint time from \cite{Stevenson_2012} using the HD149bp43 dataset ($T_\mathrm{p}=2454597.7075 \pm 0.0005$ BJD$_\mathrm{TDB}$) to the observation time of \cite{Bean_2023} using the NIRCam F444W filter ($T_\mathrm{s}=2459795.8734 \pm 0.0006$ BJD$_\mathrm{TDB}$) yields $e\cos{\omega}=-0.00164 \pm 0.000722$, consistent with the value from \cite{Winn_2008}. Assuming the $\omega$ value from \cite{Ment_2018} suggests $e \lesssim 0.007$\footnote{Secondary eclipse time from the F322W2 observation also leads to very similar result: $e \lesssim0.008$.}, which is nearly circular and thus insufficient to induce the significant red shift in $V_\mathrm{rest}$ observed in our results.

\subsection{Dependence of detection significance on $\eta$} \label{chap:param_eta}

\begin{figure}[t!]
\centering
\includegraphics[width=\linewidth]{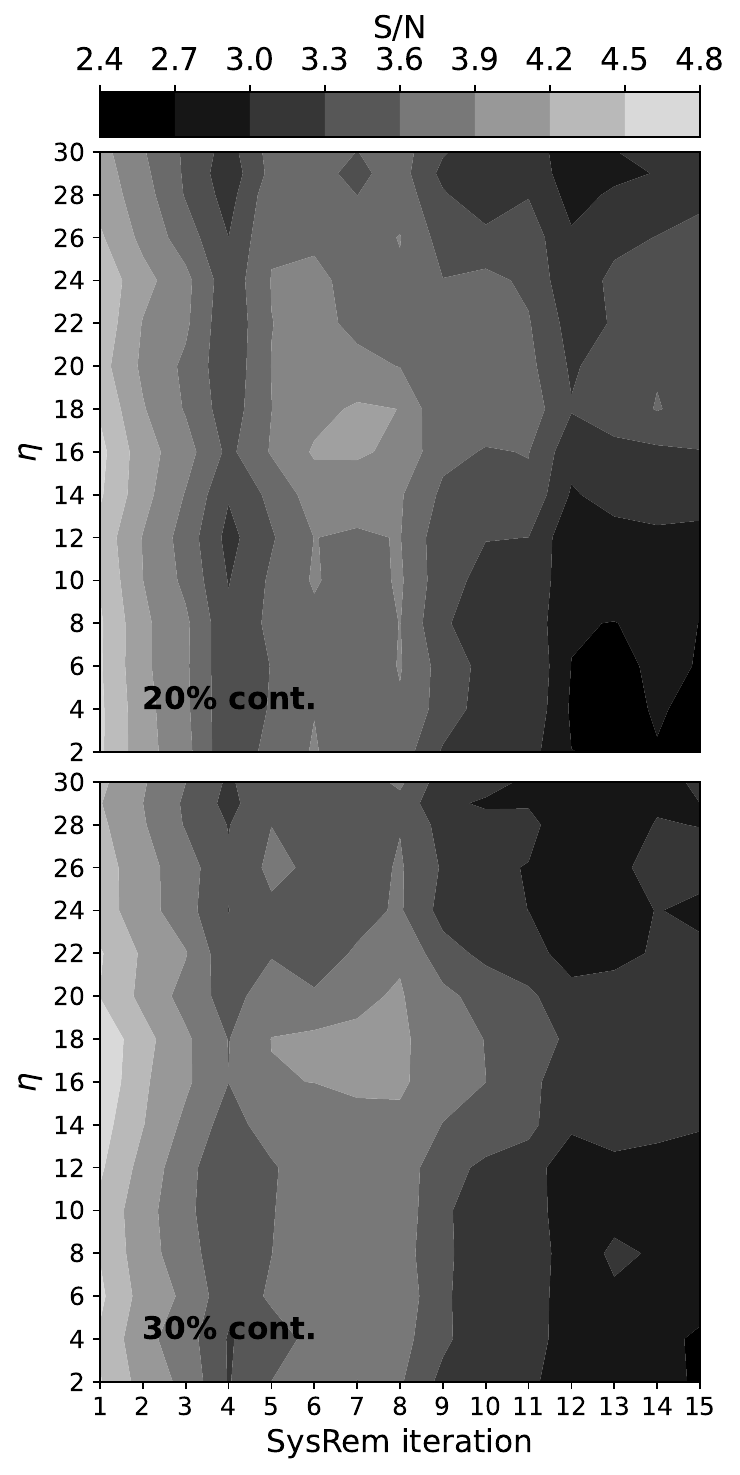}
\caption{S/N evolution with $\eta$ and SysRem iteration. S/N is taken at ($K_\mathrm{p},V_\mathrm{rest}$) $\sim$ (160.0, 2.5) $\mathrm{km\ s}^{-1}$, which is the retrieved values from S/N map in Figure \ref{fig:CEmodel_SNWelch_map}.}
\label{fig:param_eta}
\end{figure}

\cite{Cabot_2018} showed that, in such high-resolution analysis, a quantifiable parameter in the reduction steps is necessary to be able to assess how the detection significance changes as we optimize the parameter, as well as avoid bias in the results and interpretation. In this work, this parameter is $\eta$. Figure \ref{fig:param_eta} shows the detection significance in S/N at ($K_\mathrm{p},V_{\mathrm{rest}}$) $\sim$ (160.0, 2.5) $\mathrm{km\ s}^{-1}$ for different $\eta$ value (from 2 to 30 with an increment of 2) and for each SysRem iteration. Larger $\eta$ means that the data suffered from a heavier masking percentage. The lower and upper panel shows the S/N if we adopted the threshold continuum level (see Section \ref{chap:data_reduc}) at 30\% and 20\% respectively. We can see that S/N is indeed changing with different $\eta$ within the same number of iterations. This is expected because the amount of systematics (i.e., telluric contamination in this context) that will be removed by SysRem will change as $\eta$ also changes, giving different S/N within the same SysRem iteration \citep{Cabot_2018}. For iteration \#1 and 30\% continuum, the maximum S/N is achieved at $\eta=18$ (S/N = 5.0) whilst similar though slightly lower S/N values are achieved for $14\leq\eta\leq 16$. For the 20\% continuum, the maximum S/N is achieved at 4.8 for SysRem iteration \#1, and $\eta=16$ with S/N for $2\leq\eta\leq 26$ is seen to be within 0.6 difference from the maximum. We note that, at this iteration, the S/N will not fall below 4.0 for both the 20\% and 30\% continuum, for almost all $\eta$ values considered here (except for highest $\eta$ values for 20\% continuum), at least.

In addition, we can see similar trends in SysRem iteration between $\eta$ where S/N mostly decreases as the number of iterations increases (there is also, again, an unusual drop in S/N at iteration \#4 for almost all $\eta$ values as has been mentioned briefly in Section \ref{chap:evidence}). We can also see a trend of S/N values with $\eta$, although less obvious, where the highest S/N lies around $\eta=16$ and 18 within the same iteration. We note that this might not be true for other datasets as they might have different levels of systematics. In the broader context, S/N seems to be less robust with SysRem iteration than that with $\eta$. Regardless, as stated also by \cite{Cabot_2018}, we emphasize the use of quantifiable analysis parameters to be able to understand how vulnerable our detection is to these parameters and how this will affect our conclusion.

\pagebreak
\section{Summary} \label{chap:summary}

Here, using an archival CARMENES dataset, we found evidence of H$_2$O in the atmosphere of the metal-rich hot Saturn HD 149026 b at S/N $\sim4.8$ and 4.8 as well in $\alpha$-significance. We also performed a Welch-t test and confirmed the H$_2$O evidence but with a higher significance of 6.4-$\sigma$. Puzzlingly, the strength of the signal at around the same $K_\mathrm{p}, V_{\mathrm{rest}}$ recovered using HITEMP has an S/N of only $\sim2.5$. Based on the injection test, we should have detected H$_2$O with HITEMP at S/N $>4,0$. Whilst we cannot explain this discrepancy, the difference in accuracy and completeness between the two line lists surely have some effects. 

We also searched for HCN but no successful detection can be made. Additional analysis involving different C/O ratios for HCN models showed that our data is not sensitive to this species, even at a very high C/O ratio. Since this planet was indicated to have a super-solar C/O \citep{Ishizuka_2021,Bean_2023} which is indicative of a transition from oxygen- to carbon-dominated atmosphere around the relevant temperature, our result cannot rule out the possibility of HCN existence in the atmosphere.

We constrained the orbital and systemic velocity of the planet using likelihood mapping. Whilst the retrieved value for $K_\mathrm{p}$ is consistent with the expected value, $V_{\mathrm{rest}}$ is highly red-shifted to 2.57$^{+0.54}_{-0.57}$ $\mathrm{km\ s}^{-1}$. This might be explained by several scenarios such as using a circular solution on a non-eccentric orbit and/or water absorption coming from a specific red-shifting region. However, the quality of the dataset used prevents us from confirming either of these scenarios.

Finally, the relatively lower signal S/N, in addition to the aforementioned anomalous findings namely the large red-shifted $V_\mathrm{rest}$, large line width from Welch-t test, and the much lower detected S/N with HITEMP line list, emphasizes that the signal could only be regarded as an evidence of water existence, not a confirmed detection. More observations with much better data quality are indeed necessary to confirm our findings and better understand the atmosphere of this intriguing planet.

\vspace{0.8cm}

We thank the anonymous referee for the valuable feedbacks which have enhanced the quality of this manuscript. We also thank Hajime Kawahara, Yui Kawashima, John Livingston, and Ahlam Al Qasim for the helpful discussion. S.K.N is supported by JSPS KAKENHI grant No. 22K14092. M.T. is supported by JSPS KAKENHI grant No. 18H05442. L.N. acknowledges the support by the Deutsche Forschungsgemeinschaft (DFG, German Research Foundation) – Project number 314665159. A.S.L. acknowledges financial support from the Severo Ochoa grant CEX2021-001131-S funded by MCIN/AEI/ 10.13039/501100011033. The Python script used for this work can be found via the following GitHub repository: \href{https://github.com/salirafi/High-Resolution-Transmission-Spectroscopy-Analysis-of-HD-149026-b}{https://github.com/salirafi/High-Resolution-Transmission-Spectroscopy-Analysis-of-HD-149026-b}.


\bibliography{main}{}

\begin{thebibliography}{}
\expandafter\ifx\csname natexlab\endcsname\relax\def\natexlab#1{#1}\fi
\providecommand{\url}[1]{\href{#1}{#1}}
\providecommand{\dodoi}[1]{doi:~\href{http://doi.org/#1}{\nolinkurl{#1}}}
\providecommand{\doeprint}[1]{\href{http://ascl.net/#1}{\nolinkurl{http://ascl.net/#1}}}
\providecommand{\doarXiv}[1]{\href{https://arxiv.org/abs/#1}{\nolinkurl{https://arxiv.org/abs/#1}}}

\bibitem[{Alonso-Floriano {et~al.}(2019)Alonso-Floriano, S{\'{a} }nchez-L{\'{o}}pez, Snellen, L{\'{o}}pez-Puertas, Nagel, Amado, Bauer, Caballero, Czesla, Nortmann, Pall{\'{e}}, Salz, Reiners, Ribas, Quirrenbach, Aceituno, Anglada-Escud{\'{e}}, B{\'{e}}jar, Guenther, Henning, Kaminski, Kürster, Lamp{\'{o}}n, Lara, Montes, Morales, Tal-Or, Schmitt, Osorio, \& Zechmeister}]{Alonso_Floriano_2019}
Alonso-Floriano, F.~J., S{\'{a} }nchez-L{\'{o}}pez, A., Snellen, I. A.~G., {et~al.} 2019, A\&A, 621, A74, \dodoi{10.1051/0004-6361/201834339}

\bibitem[{Asplund {et~al.}(2009)Asplund, Grevesse, Sauval, \& Scott}]{Asplund_2009}
Asplund, M., Grevesse, N., Sauval, A.~J., \& Scott, P. 2009, ARA\&A, 47, 481, \dodoi{10.1146/annurev.astro.46.060407.145222}

\bibitem[{Barber {et~al.}(2014)Barber, Strange, Hill, Polyansky, Mellau, Yurchenko, \& Tennyson}]{Barber_2014}
Barber, R.~J., Strange, J.~K., Hill, C., {et~al.} 2014, \mnras, 437, 1828, \dodoi{10.1093/mnras/stt2011}

\bibitem[{Barber {et~al.}(2006)Barber, Tennyson, Harris, \& Tolchenov}]{Barber_2006}
Barber, R.~J., Tennyson, J., Harris, G.~J., \& Tolchenov, R.~N. 2006, \mnras, 368, 1087, \dodoi{10.1111/j.1365-2966.2006.10184.x}

\bibitem[{Bean {et~al.}(2023)Bean, Xue, August, Lunine, Zhang, Thorngren, Tsai, Stassun, Schlawin, Ahrer, Ih, \& Mansfield}]{Bean_2023}
Bean, J.~L., Xue, Q., August, P.~C., {et~al.} 2023, Nature, 618, 43, \dodoi{10.1038/s41586-023-05984-y}

\bibitem[{Birkby(2018)}]{Birkby_2018}
Birkby, J.~L. 2018, Exoplanet Atmospheres at High Spectral Resolution,  arXiv, \dodoi{10.48550/ARXIV.1806.04617}

\bibitem[{Birkby {et~al.}(2017)Birkby, de~Kok, Brogi, Schwarz, \& Snellen}]{Birkby_2017}
Birkby, J.~L., de~Kok, R.~J., Brogi, M., Schwarz, H., \& Snellen, I. A.~G. 2017, \aj, 153, 138, \dodoi{10.3847/1538-3881/aa5c87}

\bibitem[{Blain {et~al.}(2024)Blain, Sánchez-López, \& Mollière}]{Blain_2024}
Blain, D., Sánchez-López, A., \& Mollière, P. 2024, \aj, 167, 179, \dodoi{10.3847/1538-3881/ad2c8b}

\bibitem[{{Bonomo} {et~al.}(2017){Bonomo}, {Desidera}, {Benatti}, {Borsa}, {Crespi}, {Damasso}, {Lanza}, {Sozzetti}, {Lodato}, {Marzari}, {Boccato}, {Claudi}, {Cosentino}, {Covino}, {Gratton}, {Maggio}, {Micela}, {Molinari}, {Pagano}, {Piotto}, {Poretti}, {Smareglia}, {Affer}, {Biazzo}, {Bignamini}, {Esposito}, {Giacobbe}, {H{\'e}brard}, {Malavolta}, {Maldonado}, {Mancini}, {Martinez Fiorenzano}, {Masiero}, {Nascimbeni}, {Pedani}, {Rainer}, \& {Scandariato}}]{Bonomo_2017}
{Bonomo}, A.~S., {Desidera}, S., {Benatti}, S., {et~al.} 2017, A\&A, 602, A107, \dodoi{10.1051/0004-6361/201629882}

\bibitem[{Borsa {et~al.}(2021)Borsa, {Allart, R.}, {Casasayas-Barris, N.}, {Tabernero, H.}, {Zapatero Osorio, M. R.}, {Cristiani, S.}, {Pepe, F.}, {Rebolo, R.}, {Santos, N. C.}, {Adibekyan, V.}, {Bourrier, V.}, {Demangeon, O. D. S.}, {Ehrenreich, D.}, {Pall\'e, E.}, {Sousa, S.}, {Lillo-Box, J.}, {Lovis, C.}, {Micela, G.}, {Oshagh, M.}, {Poretti, E.}, {Sozzetti, A.}, {Allende Prieto, C.}, {Alibert, Y.}, {Amate, M.}, {Benz, W.}, {Bouchy, F.}, {Cabral, A.}, {Dekker, H.}, {D\'{}Odorico, V.}, {Di Marcantonio, P.}, {Figueira, P.}, {Genova Santos, R.}, {Gonz\'alez Hern\'andez, J. I.}, {Lo Curto, G.}, {Manescau, A.}, {Martins, C. J. A. P.}, {M\'egevand, D.}, {Mehner, A.}, {Molaro, P.}, {Nunes, N. J.}, {Riva, M.}, {Su\'arez Mascare\~no, A.}, {Udry, S.}, \& {Zerbi, F.}}]{Borsa_2021}
Borsa, F., {Allart, R.}, {Casasayas-Barris, N.}, {et~al.} 2021, A\&A, 645, A24, \dodoi{10.1051/0004-6361/202039344}

\bibitem[{Brewer {et~al.}(2017)Brewer, Fischer, \& Madhusudhan}]{Brewer_2017}
Brewer, J.~M., Fischer, D.~A., \& Madhusudhan, N. 2017, \aj, 153, 83, \dodoi{10.3847/1538-3881/153/2/83}

\bibitem[{Broeg \& Wuchterl(2007)}]{Broeg_2007}
Broeg, C., \& Wuchterl, G. 2007, \mnras: Letters, 376, L62, \dodoi{10.1111/j.1745-3933.2007.00287.x}

\bibitem[{Brogi \& Line(2019)}]{Brogi_2019}
Brogi, M., \& Line, M.~R. 2019, \aj, 157, 114, \dodoi{10.3847/1538-3881/aaffd3}

\bibitem[{Brogi {et~al.}(2023)Brogi, Emeka-Okafor, Line, Gandhi, Pino, Kempton, Rauscher, Parmentier, Bean, Mace, Cowan, Shkolnik, Wardenier, Mansfield, Welbanks, Smith, Fortney, Birkby, Zalesky, Dang, Patience, \& Désert}]{Brogi_2023}
Brogi, M., Emeka-Okafor, V., Line, M.~R., {et~al.} 2023, \aj, 165, 91, \dodoi{10.3847/1538-3881/acaf5c}

\bibitem[{Burrows {et~al.}(2007)Burrows, Hubeny, Budaj, \& Hubbard}]{Burrows_2007}
Burrows, A., Hubeny, I., Budaj, J., \& Hubbard, W.~B. 2007, \apj, 661, 502, \dodoi{10.1086/514326}

\bibitem[{{Caballero} {et~al.}(2016){Caballero}, {Gu{\`a}rdia}, {L{\'o}pez del Fresno}, {Zechmeister}, {de Juan}, {Alonso-Floriano}, {Amado}, {Colom{\'e}}, {Cort{\'e}s-Contreras}, {Garc{\'\i}a-Piquer}, {Gesa}, {de Guindos}, {Hagen}, {Helmling}, {Hern{\'a}ndez Casta{\~n}o}, {K{\"u}rster}, {L{\'o}pez-Santiago}, {Montes}, {Morales Mu{\~n}oz}, {Pavlov}, {Quirrenbach}, {Reiners}, {Ribas}, {Seifert}, \& {Solano}}]{Caballero_2016}
{Caballero}, J.~A., {Gu{\`a}rdia}, J., {L{\'o}pez del Fresno}, M., {et~al.} 2016, in Society of Photo-Optical Instrumentation Engineers (SPIE) Conference Series, Vol. 9910, Observatory Operations: Strategies, Processes, and Systems VI, ed. A.~B. {Peck}, R.~L. {Seaman}, \& C.~R. {Benn}, 99100E, \dodoi{10.1117/12.2233574}

\bibitem[{Cabot {et~al.}(2018)Cabot, Madhusudhan, Hawker, \& Gandhi}]{Cabot_2018}
Cabot, S. H.~C., Madhusudhan, N., Hawker, G.~A., \& Gandhi, S. 2018, \mnras, 482, 4422, \dodoi{10.1093/mnras/sty2994}

\bibitem[{Charbonneau {et~al.}(2002)Charbonneau, Brown, Noyes, \& Gilliland}]{Charbonneau_2002}
Charbonneau, D., Brown, T.~M., Noyes, R.~W., \& Gilliland, R.~L. 2002, \apj, 568, 377, \dodoi{10.1086/338770}

\bibitem[{{Cont} {et~al.}(2021){Cont}, {Yan}, {Reiners}, {Casasayas-Barris}, {Molli{\`e}re}, {Pall{\'e}}, {Henning}, {Nortmann}, {Stangret}, {Czesla}, {L{\'o}pez-Puertas}, {S{\'a}nchez-L{\'o}pez}, {Rodler}, {Ribas}, {Quirrenbach}, {Caballero}, {Amado}, {Carone}, {Khaimova}, {Kreidberg}, {Molaverdikhani}, {Montes}, {Morello}, {Nagel}, {Oshagh}, \& {Zechmeister}}]{Cont2021}
{Cont}, D., {Yan}, F., {Reiners}, A., {et~al.} 2021, A\&A, 651, A33, \dodoi{10.1051/0004-6361/202140732}

\bibitem[{Cridland {et~al.}(2019)Cridland, {Eistrup, Christian}, \& {van Dishoeck, Ewine F.}}]{Cridland_2019}
Cridland, A.~J., {Eistrup, Christian}, \& {van Dishoeck, Ewine F.} 2019, A\&A, 627, A127, \dodoi{10.1051/0004-6361/201834378}

\bibitem[{Deibert {et~al.}(2021)Deibert, de~Mooij, Jayawardhana, Ridden-Harper, Sivanandam, Karjalainen, \& Karjalainen}]{Deibert_2021}
Deibert, E.~K., de~Mooij, E. J.~W., Jayawardhana, R., {et~al.} 2021, \aj, 161, 209, \dodoi{10.3847/1538-3881/abe768}

\bibitem[{Deming {et~al.}(2006)Deming, Harrington, Seager, \& Richardson}]{Deming_2006}
Deming, D., Harrington, J., Seager, S., \& Richardson, L.~J. 2006, \apj, 644, 560, \dodoi{10.1086/503358}

\bibitem[{Dodson-Robinson \& Bodenheimer(2009)}]{Dodson_Robinson_2009}
Dodson-Robinson, S.~E., \& Bodenheimer, P. 2009, \apj, 695, L159, \dodoi{10.1088/0004-637x/695/2/l159}

\bibitem[{Flowers {et~al.}(2019)Flowers, Brogi, Rauscher, Kempton, \& Chiavassa}]{Flowers_2019}
Flowers, E., Brogi, M., Rauscher, E., Kempton, E. M.-R., \& Chiavassa, A. 2019, \aj, 157, 209, \dodoi{10.3847/1538-3881/ab164c}

\bibitem[{Fortney(2005)}]{Fortney_2005}
Fortney, J.~J. 2005, \mnras, 364, 649, \dodoi{10.1111/j.1365-2966.2005.09587.x}

\bibitem[{Fortney {et~al.}(2013)Fortney, Mordasini, Nettelmann, Kempton, Greene, \& Zahnle}]{Fortney_2013}
Fortney, J.~J., Mordasini, C., Nettelmann, N., {et~al.} 2013, \apj, 775, 80, \dodoi{10.1088/0004-637X/775/1/80}

\bibitem[{Fortney {et~al.}(2006)Fortney, Saumon, Marley, Lodders, \& Freedman}]{Fortney_2006}
Fortney, J.~J., Saumon, D., Marley, M.~S., Lodders, K., \& Freedman, R.~S. 2006, \apj, 642, 495, \dodoi{10.1086/500920}

\bibitem[{Fortney {et~al.}(2010)Fortney, Shabram, Showman, Lian, Freedman, Marley, \& Lewis}]{Fortney_2010}
Fortney, J.~J., Shabram, M., Showman, A.~P., {et~al.} 2010, \apj, 709, 1396, \dodoi{10.1088/0004-637X/709/2/1396}

\bibitem[{Gagnebin {et~al.}(2024)Gagnebin, Mukherjee, Fortney, \& Batalha}]{Gagnebin_2024}
Gagnebin, A., Mukherjee, S., Fortney, J.~J., \& Batalha, N.~E. 2024, The atmosphere of HD 149026b: Low metal-enrichment and weak energy transport.
\newblock \doarXiv{2404.17658}

\bibitem[{{Gaia Collaboration}(2018)}]{Gaia_2018}
{Gaia Collaboration}. 2018, VizieR Online Data Catalog, I/345

\bibitem[{{Gaia Collaboration}(2020)}]{Gaia_2020}
---. 2020, VizieR Online Data Catalog, I/350

\bibitem[{Gandhi {et~al.}(2020)Gandhi, Brogi, Yurchenko, Tennyson, Coles, Webb, Birkby, Guilluy, Hawker, Madhusudhan, Bonomo, \& Sozzetti}]{Gandhi_2020}
Gandhi, S., Brogi, M., Yurchenko, S.~N., {et~al.} 2020, \mnras, 495, 224, \dodoi{10.1093/mnras/staa981}

\bibitem[{Giacobbe {et~al.}(2021)Giacobbe, Brogi, Gandhi, Cubillos, Bonomo, Sozzetti, Fossati, Guilluy, Carleo, Rainer, Harutyunyan, Borsa, Pino, Nascimbeni, Benatti, Biazzo, Bignamini, Chubb, Claudi, Cosentino, Covino, Damasso, Desidera, Fiorenzano, Ghedina, Lanza, Leto, Maggio, Malavolta, Maldonado, Micela, Molinari, Pagano, Pedani, Piotto, Poretti, Scandariato, Yurchenko, Fantinel, Galli, Lodi, Sanna, \& Tozzi}]{Giacobbe_2021}
Giacobbe, P., Brogi, M., Gandhi, S., {et~al.} 2021, Nature, 592, 205, \dodoi{10.1038/s41586-021-03381-x}

\bibitem[{Gibson {et~al.}(2018)Gibson, de~Mooij, Evans, Merritt, Nikolov, Sing, \& Watson}]{Gibson_2018}
Gibson, N.~P., de~Mooij, E. J.~W., Evans, T.~M., {et~al.} 2018, \mnras, 482, 606, \dodoi{10.1093/mnras/sty2722}

\bibitem[{Gibson {et~al.}(2022)Gibson, Nugroho, Lothringer, Maguire, \& Sing}]{Gibson_2022}
Gibson, N.~P., Nugroho, S.~K., Lothringer, J., Maguire, C., \& Sing, D.~K. 2022, \mnras, 512, 4618, \dodoi{10.1093/mnras/stac091}

\bibitem[{Gibson {et~al.}(2020)Gibson, Merritt, Nugroho, Cubillos, de~Mooij, Mikal-Evans, Fossati, Lothringer, Nikolov, Sing, Spake, Watson, \& Wilson}]{Gibson_2020}
Gibson, N.~P., Merritt, S., Nugroho, S.~K., {et~al.} 2020, \mnras, 493, 2215, \dodoi{10.1093/mnras/staa228}

\bibitem[{{Grimm} \& {Heng}(2015)}]{grimm2015}
{Grimm}, S.~L., \& {Heng}, K. 2015, ApJ, 808, 182, \dodoi{10.1088/0004-637X/808/2/182}

\bibitem[{Grimm {et~al.}(2021)Grimm, Malik, Kitzmann, Guzmán-Mesa, Hoeijmakers, Fisher, Mendonça, Yurchenko, Tennyson, Alesina, Buchschacher, Burnier, Segransan, Kurucz, \& Heng}]{Grimm_2021}
Grimm, S.~L., Malik, M., Kitzmann, D., {et~al.} 2021, \apjs, 253, 30, \dodoi{10.3847/1538-4365/abd773}

\bibitem[{Guillot(2010)}]{Guillot_2010}
Guillot, T. 2010, A\&A, 520, A27, \dodoi{10.1051/0004-6361/200913396}

\bibitem[{{Guilluy} {et~al.}(2019){Guilluy}, {Sozzetti}, {Brogi}, {Bonomo}, {Giacobbe}, {Claudi}, \& {Benatti}}]{Guilluy_2019}
{Guilluy}, G., {Sozzetti}, A., {Brogi}, M., {et~al.} 2019, A\&A, 625, A107, \dodoi{10.1051/0004-6361/201834615}

\bibitem[{Harris {et~al.}(2006)Harris, Tennyson, Kaminsky, Pavlenko, \& Jones}]{Harris_2006}
Harris, G.~J., Tennyson, J., Kaminsky, B.~M., Pavlenko, Y.~V., \& Jones, H. R.~A. 2006, \mnras, 367, 400, \dodoi{10.1111/j.1365-2966.2005.09960.x}

\bibitem[{Herman {et~al.}(2020)Herman, de~Mooij, Jayawardhana, \& Brogi}]{Herman_2020}
Herman, M.~K., de~Mooij, E. J.~W., Jayawardhana, R., \& Brogi, M. 2020, \aj, 160, 93, \dodoi{10.3847/1538-3881/ab9e77}

\bibitem[{Herman {et~al.}(2022)Herman, de~Mooij, Nugroho, Gibson, \& Jayawardhana}]{Herman_2022}
Herman, M.~K., de~Mooij, E. J.~W., Nugroho, S.~K., Gibson, N.~P., \& Jayawardhana, R. 2022, \aj, 163, 248, \dodoi{10.3847/1538-3881/ac5f4d}

\bibitem[{Hoeijmakers {et~al.}(2019)Hoeijmakers, Ehrenreich, Kitzmann, Allart, Grimm, Seidel, Wyttenbach, Pino, Nielsen, Fisher, Rimmer, Bourrier, Cegla, Lavie, Lovis, Patzer, Stock, Pepe, \& Heng}]{Hoeijmakers_2019}
Hoeijmakers, H.~J., Ehrenreich, D., Kitzmann, D., {et~al.} 2019, A\&A, 627, A165, \dodoi{10.1051/0004-6361/201935089}

\bibitem[{Ikoma {et~al.}(2006)Ikoma, Guillot, Genda, Tanigawa, \& Ida}]{Ikoma_2006}
Ikoma, M., Guillot, T., Genda, H., Tanigawa, T., \& Ida, S. 2006, \apj, 650, 1150, \dodoi{10.1086/507088}

\bibitem[{Ishizuka {et~al.}(2021)Ishizuka, Kawahara, Nugroho, Kawashima, Hirano, \& Tamura}]{Ishizuka_2021}
Ishizuka, M., Kawahara, H., Nugroho, S.~K., {et~al.} 2021, \aj, 161, 153, \dodoi{10.3847/1538-3881/abdb25}

\bibitem[{{Kang} {et~al.}(2011){Kang}, {Lee}, \& {Kim}}]{Kang_2011}
{Kang}, W., {Lee}, S.-G., \& {Kim}, K.-M. 2011, \apj, 736, 87, \dodoi{10.1088/0004-637X/736/2/87}

\bibitem[{Kawahara {et~al.}(2022)Kawahara, Kawashima, Masuda, Crossfield, Pannier, \& van~den Bekerom}]{Kawahara_2022}
Kawahara, H., Kawashima, Y., Masuda, K., {et~al.} 2022, \apjs, 258, 31, \dodoi{10.3847/1538-4365/ac3b4d}

\bibitem[{Kesseli {et~al.}(2022)Kesseli, Snellen, Casasayas-Barris, Mollière, \& Sánchez-López}]{Kesseli_2022}
Kesseli, A.~Y., Snellen, I. A.~G., Casasayas-Barris, N., Mollière, P., \& Sánchez-López, A. 2022, \aj, 163, 107, \dodoi{10.3847/1538-3881/ac4336}

\bibitem[{Knutson {et~al.}(2009)Knutson, Charbonneau, Cowan, Fortney, Showman, Agol, \& Henry}]{Knutson_2009}
Knutson, H.~A., Charbonneau, D., Cowan, N.~B., {et~al.} 2009, \apj, 703, 769, \dodoi{10.1088/0004-637X/703/1/769}

\bibitem[{Knutson {et~al.}(2014)Knutson, Fulton, Montet, Kao, Ngo, Howard, Crepp, Hinkley, Bakos, Batygin, Johnson, Morton, \& Muirhead}]{Knutson_2014}
Knutson, H.~A., Fulton, B.~J., Montet, B.~T., {et~al.} 2014, \apj, 785, 126, \dodoi{10.1088/0004-637X/785/2/126}

\bibitem[{Kreidberg(2015)}]{Kreidberg_2015}
Kreidberg, L. 2015, \pasp, 127, 1161, \dodoi{10.1086/683602}

\bibitem[{{Landman} {et~al.}(2021){Landman}, {S{\'a}nchez-L{\'o}pez}, {Molli{\`e}re}, {Kesseli}, {Louca}, \& {Snellen}}]{Landman_2021}
{Landman}, R., {S{\'a}nchez-L{\'o}pez}, A., {Molli{\`e}re}, P., {et~al.} 2021, A\&A, 656, A119, \dodoi{10.1051/0004-6361/202141696}

\bibitem[{Line {et~al.}(2021)Line, Brogi, Bean, Gandhi, Zalesky, Parmentier, Smith, Mace, Mansfield, Kempton, Fortney, Shkolnik, Patience, Rauscher, D{\'e}sert, \& Wardenier}]{Line_2021}
Line, M.~R., Brogi, M., Bean, J.~L., {et~al.} 2021, Nature, 598, 580, \dodoi{10.1038/s41586-021-03912-6}

\bibitem[{Machalek {et~al.}(2009)Machalek, McCullough, Burrows, Burke, Hora, \& Johns-Krull}]{Machalek_2009}
Machalek, P., McCullough, P.~R., Burrows, A., {et~al.} 2009, \apj, 701, 514, \dodoi{10.1088/0004-637X/701/1/514}

\bibitem[{Madhusudhan(2012)}]{Madhusudhan_2012}
Madhusudhan, N. 2012, \apj, 758, 36, \dodoi{10.1088/0004-637x/758/1/36}

\bibitem[{Madhusudhan {et~al.}(2014)Madhusudhan, Amin, \& Kennedy}]{Madhusudhan_2014}
Madhusudhan, N., Amin, M.~A., \& Kennedy, G.~M. 2014, \apj, 794, L12, \dodoi{10.1088/2041-8205/794/1/l12}

\bibitem[{{Matsuo} {et~al.}(2007){Matsuo}, {Shibai}, {Ootsubo}, \& {Tamura}}]{Matsuo_2007}
{Matsuo}, T., {Shibai}, H., {Ootsubo}, T., \& {Tamura}, M. 2007, \apj, 662, 1282, \dodoi{10.1086/517964}

\bibitem[{{Ment} {et~al.}(2018){Ment}, {Fischer}, {Bakos}, {Howard}, \& {Isaacson}}]{Ment_2018}
{Ment}, K., {Fischer}, D.~A., {Bakos}, G., {Howard}, A.~W., \& {Isaacson}, H. 2018, \aj, 156, 213, \dodoi{10.3847/1538-3881/aae1f5}

\bibitem[{Merritt {et~al.}(2020)Merritt, Gibson, Nugroho, de~Mooij, Hooton, Matthews, McKemmish, Mikal-Evans, Nikolov, Sing, Spake, \& Watson}]{Merritt_2020}
Merritt, S.~R., Gibson, N.~P., Nugroho, S.~K., {et~al.} 2020, A\&A, 636, A117, \dodoi{10.1051/0004-6361/201937409}

\bibitem[{Mollière {et~al.}(2015)Mollière, van Boekel, Dullemond, Henning, \& Mordasini}]{Molliere_2015}
Mollière, P., van Boekel, R., Dullemond, C., Henning, T., \& Mordasini, C. 2015, \apj, 813, 47, \dodoi{10.1088/0004-637X/813/1/47}

\bibitem[{Mollière {et~al.}(2019)Mollière, Wardenier, van Boekel, Henning, Molaverdikhani, \& Snellen}]{Molliere_2019}
Mollière, P., Wardenier, J.~P., van Boekel, R., {et~al.} 2019, A\&A, 627, A67, \dodoi{10.1051/0004-6361/201935470}

\bibitem[{{Montalto, M.} {et~al.}(2011){Montalto, M.}, {Santos, N. C.}, {Boisse, I.}, {Boué, G.}, {Figueira, P.}, \& {Sousa, S.}}]{Montalto_2011}
{Montalto, M.}, {Santos, N. C.}, {Boisse, I.}, {et~al.} 2011, A\&A, 528, L17, \dodoi{10.1051/0004-6361/201116492}

\bibitem[{Mordasini {et~al.}(2016)Mordasini, van Boekel, Mollière, Henning, \& Benneke}]{Mordasini_2016}
Mordasini, C., van Boekel, R., Mollière, P., Henning, T., \& Benneke, B. 2016, \apj, 832, 41, \dodoi{10.3847/0004-637X/832/1/41}

\bibitem[{Moses {et~al.}(2013)Moses, Line, Visscher, Richardson, Nettelmann, Fortney, Barman, Stevenson, \& Madhusudhan}]{Moses_2013}
Moses, J.~I., Line, M.~R., Visscher, C., {et~al.} 2013, \apj, 777, 34, \dodoi{10.1088/0004-637X/777/1/34}

\bibitem[{Nugroho {et~al.}(2020{\natexlab{a}})Nugroho, Gibson, de~Mooij, Herman, Watson, Kawahara, \& Merritt}]{Nugroho_2020b}
Nugroho, S.~K., Gibson, N.~P., de~Mooij, E. J.~W., {et~al.} 2020{\natexlab{a}}, \apj, 898, L31, \dodoi{10.3847/2041-8213/aba4b6}

\bibitem[{Nugroho {et~al.}(2020{\natexlab{b}})Nugroho, Gibson, de~Mooij, Watson, Kawahara, \& Merritt}]{Nugroho_2020a}
---. 2020{\natexlab{b}}, \mnras, 496, 504, \dodoi{10.1093/mnras/staa1459}

\bibitem[{Nugroho {et~al.}(2017)Nugroho, Kawahara, Masuda, Hirano, Kotani, \& Tajitsu}]{Nugroho_2017}
Nugroho, S.~K., Kawahara, H., Masuda, K., {et~al.} 2017, \aj, 154, 221, \dodoi{10.3847/1538-3881/aa9433}

\bibitem[{Nugroho {et~al.}(2021)Nugroho, Kawahara, Gibson, de~Mooij, Hirano, Kotani, Kawashima, Masuda, Brogi, Birkby, Watson, Tamura, Zwintz, Harakawa, Kudo, Kuzuhara, Hodapp, Ishizuka, Jacobson, Konishi, Kurokawa, Nishikawa, Omiya, Serizawa, Ueda, \& Vievard}]{Nugroho_2021}
Nugroho, S.~K., Kawahara, H., Gibson, N.~P., {et~al.} 2021, \apjl, 910, L9, \dodoi{10.3847/2041-8213/abec71}

\bibitem[{Oliva {et~al.}(2015)Oliva, Origlia, Scuderi, Benatti, Carleo, Lapenna, Mucciarelli, Baffa, Biliotti, Carbonaro, {et~al.}}]{Oliva_2015}
Oliva, E., Origlia, L., Scuderi, S., {et~al.} 2015, A\&A, 581, A47

\bibitem[{Pelletier {et~al.}(2021)Pelletier, Benneke, Darveau-Bernier, Boucher, Cook, Piaulet, Coulombe, Étienne Artigau, Lafrenière, Delisle, Allart, Doyon, Donati, Fouqué, Moutou, Cadieux, Delfosse, Hébrard, Martins, Martioli, \& Vandal}]{Pelletier_2021}
Pelletier, S., Benneke, B., Darveau-Bernier, A., {et~al.} 2021, \aj, 162, 73, \dodoi{10.3847/1538-3881/ac0428}

\bibitem[{Pelletier {et~al.}(2023)Pelletier, Benneke, Ali-Dib, Prinoth, Kasper, Seifahrt, Bean, Debras, Klein, Bazinet, Hoeijmakers, Kesseli, Lim, Carmona, Pino, Casasayas-Barris, Hood, \& St{\"u}rmer}]{Pelletier_2023}
Pelletier, S., Benneke, B., Ali-Dib, M., {et~al.} 2023, Nature, 619, 491, \dodoi{10.1038/s41586-023-06134-0}

\bibitem[{Polyansky {et~al.}(2018)Polyansky, Kyuberis, Zobov, Tennyson, Yurchenko, \& Lodi}]{Polyansky_2018}
Polyansky, O.~L., Kyuberis, A.~A., Zobov, N.~F., {et~al.} 2018, \mnras, 480, 2597, \dodoi{10.1093/mnras/sty1877}

\bibitem[{{Prinoth} {et~al.}(2022){Prinoth}, {Hoeijmakers}, {Kitzmann}, {Sandvik}, {Seidel}, {Lendl}, {Borsato}, {Thorsbro}, {Anderson}, {Barrado}, {Kravchenko}, {Allart}, {Bourrier}, {Cegla}, {Ehrenreich}, {Fisher}, {Lovis}, {Guzm{\'a}n-Mesa}, {Grimm}, {Hooton}, {Morris}, {Oreshenko}, {Pino}, \& {Heng}}]{Prinoth_2022}
{Prinoth}, B., {Hoeijmakers}, H.~J., {Kitzmann}, D., {et~al.} 2022, Nature Astronomy, 6, 449, \dodoi{10.1038/s41550-021-01581-z}

\bibitem[{{Ramkumar} {et~al.}(2023){Ramkumar}, {Gibson}, {Nugroho}, {Maguire}, \& {Fortune}}]{Ramkumar2023}
{Ramkumar}, S., {Gibson}, N.~P., {Nugroho}, S.~K., {Maguire}, C., \& {Fortune}, M. 2023, \mnras, 525, 2985, \dodoi{10.1093/mnras/stad2476}

\bibitem[{Ridden-Harper {et~al.}(2023)Ridden-Harper, Nugroho, Flagg, Jayawardhana, Turner, de~Mooij, MacDonald, Deibert, Tamura, Kotani, Hirano, Kuzuhara, Omiya, \& Kusakabe}]{Ridden-Harper_2023}
Ridden-Harper, A., Nugroho, S.~K., Flagg, L., {et~al.} 2023, \aj, 165, 170, \dodoi{10.3847/1538-3881/acbd39}

\bibitem[{{Rothman} {et~al.}(2010){Rothman}, {Gordon}, {Barber}, {Dothe}, {Gamache}, {Goldman}, {Perevalov}, {Tashkun}, \& {Tennyson}}]{Rothman_2010}
{Rothman}, L.~S., {Gordon}, I.~E., {Barber}, R.~J., {et~al.} 2010, \jqsrt, 111, 2139, \dodoi{10.1016/j.jqsrt.2010.05.001}

\bibitem[{{Rousselot} {et~al.}(2000){Rousselot}, {Lidman}, {Cuby}, {Moreels}, \& {Monnet}}]{Rousselot2000}
{Rousselot}, P., {Lidman}, C., {Cuby}, J.~G., {Moreels}, G., \& {Monnet}, G. 2000, A\&A, 354, 1134

\bibitem[{S\'anchez-L\'opez {et~al.}(2022)S\'anchez-L\'opez, Landman, Molli\`ere, Casasayas-Barris, Kesseli, \& Snellen}]{Sanchez-Lopez_2022}
S\'anchez-L\'opez, A., Landman, R., Molli\`ere, P., {et~al.} 2022, A\&A, 661, A78, \dodoi{10.1051/0004-6361/202142591}

\bibitem[{S\'anchez-L\'opez {et~al.}(2019)S\'anchez-L\'opez, Alonso-Floriano, L\'opez-Puertas, Snellen, Funke, Nagel, Bauer, Amado, Caballero, Czesla, Nortmann, Pall\'e, Salz, Reiners, Ribas, Quirrenbach, Anglada-Escud\'e, B\'ejar, Casasayas-Barris, Galad\'{\i}-Enr\'{\i}quez, Guenther, Henning, Kaminski, K\"urster, Lamp\'on, Lara, Montes, Morales, Stangret, Tal-Or, Sanz-Forcada, Schmitt, Zapatero~Osorio, \& Zechmeister}]{Sanchez-Lopez_2019}
S\'anchez-L\'opez, A., Alonso-Floriano, F.~J., L\'opez-Puertas, M., {et~al.} 2019, A\&A, 630, A53, \dodoi{10.1051/0004-6361/201936084}

\bibitem[{Sato {et~al.}(2005)Sato, Fischer, Henry, Laughlin, Butler, Marcy, Vogt, Bodenheimer, Ida, Toyota, Wolf, Valenti, Boyd, Johnson, Wright, Ammons, Robinson, Strader, McCarthy, Tah, \& Minniti}]{Sato_2005}
Sato, B., Fischer, D.~A., Henry, G.~W., {et~al.} 2005, \apj, 633, 465, \dodoi{10.1086/449306}

\bibitem[{Schneider \& Bitsch(2021)}]{Schneider_2021}
Schneider, A.~D., \& Bitsch, B. 2021, A\&A, 654, A71

\bibitem[{Seager \& Sasselov(2000)}]{Seager_2000}
Seager, S., \& Sasselov, D.~D. 2000, \apj, 537, 916, \dodoi{10.1086/309088}

\bibitem[{Showman {et~al.}(2012)Showman, Fortney, Lewis, \& Shabram}]{Showman_2013}
Showman, A.~P., Fortney, J.~J., Lewis, N.~K., \& Shabram, M. 2012, \apj, 762, 24, \dodoi{10.1088/0004-637X/762/1/24}

\bibitem[{Snellen {et~al.}(2010)Snellen, De~Kok, De~Mooij, \& Albrecht}]{Snellen_2010}
Snellen, I.~A., De~Kok, R.~J., De~Mooij, E.~J., \& Albrecht, S. 2010, Nature, 465, 1049

\bibitem[{Stangret {et~al.}(2022)Stangret, {Casasayas-Barris, N.}, {Pall\'e, E.}, {Orell-Miquel, J.}, {Morello, G.}, {Luque, R.}, {Nowak, G.}, \& {Yan, F.}}]{Stangret_2022}
Stangret, M., {Casasayas-Barris, N.}, {Pall\'e, E.}, {et~al.} 2022, A\&A, 662, A101, \dodoi{10.1051/0004-6361/202141799}

\bibitem[{{Stassun} {et~al.}(2017){Stassun}, {Collins}, \& {Gaudi}}]{Stassun_2017}
{Stassun}, K.~G., {Collins}, K.~A., \& {Gaudi}, B.~S. 2017, \apj, 153, 136, \dodoi{10.3847/1538-3881/aa5df3}

\bibitem[{Stevenson {et~al.}(2012)Stevenson, Harrington, Fortney, Loredo, Hardy, Nymeyer, Bowman, Cubillos, Bowman, \& Hardin}]{Stevenson_2012}
Stevenson, K.~B., Harrington, J., Fortney, J.~J., {et~al.} 2012, \apj, 754, 136, \dodoi{10.1088/0004-637x/754/2/136}

\bibitem[{Stock {et~al.}(2018)Stock, Kitzmann, Patzer, \& Sedlmayr}]{Stock_2018}
Stock, J.~W., Kitzmann, D., Patzer, A. B.~C., \& Sedlmayr, E. 2018, \mnras, \dodoi{10.1093/mnras/sty1531}

\bibitem[{Tabernero {et~al.}(2012)Tabernero, Montes, \& Gonz\'alez~Hern\'andez}]{Tabernero_2012}
Tabernero, H.~M., Montes, D., \& Gonz\'alez~Hern\'andez, J.~I. 2012, A\&A, 547, A13, \dodoi{10.1051/0004-6361/201117506}

\bibitem[{{Tamuz} {et~al.}(2005){Tamuz}, {Mazeh}, \& {Zucker}}]{Tamuz_2005}
{Tamuz}, O., {Mazeh}, T., \& {Zucker}, S. 2005, \mnras, 356, 1466, \dodoi{10.1111/j.1365-2966.2004.08585.x}

\bibitem[{{Tinetti} {et~al.}(2012){Tinetti}, {Tennyson}, {Griffith}, \& {Waldmann}}]{Tinetti_2012}
{Tinetti}, G., {Tennyson}, J., {Griffith}, C.~A., \& {Waldmann}, I. 2012, Philosophical Transactions of the Royal Society of London Series A, 370, 2749, \dodoi{10.1098/rsta.2011.0338}

\bibitem[{Tinetti {et~al.}(2007)Tinetti, Vidal-Madjar, Liang, Beaulieu, Yung, Carey, Barber, Tennyson, Ribas, Allard, {et~al.}}]{Tinetti_2007}
Tinetti, G., Vidal-Madjar, A., Liang, M.-C., {et~al.} 2007, Nature, 448, 169

\bibitem[{van Sluijs {et~al.}(2023)van Sluijs, Birkby, Lothringer, Lee, Crossfield, Parmentier, Brogi, Kulesa, McCarthy, \& Charbonneau}]{vanSluijs_2022}
van Sluijs, L., Birkby, J.~L., Lothringer, J., {et~al.} 2023, \mnras, 522, 2145, \dodoi{10.1093/mnras/stad1103}

\bibitem[{Wardenier {et~al.}(2021)Wardenier, Parmentier, Lee, Line, \& Gharib-Nezhad}]{Wardenier_2021}
Wardenier, J.~P., Parmentier, V., Lee, E. K.~H., Line, M.~R., \& Gharib-Nezhad, E. 2021, \mnras, 506, 1258, \dodoi{10.1093/mnras/stab1797}

\bibitem[{Webb {et~al.}(2020)Webb, Brogi, Gandhi, Line, Birkby, Chubb, Snellen, \& Yurchenko}]{Webb_2020}
Webb, R.~K., Brogi, M., Gandhi, S., {et~al.} 2020, \mnras, 494, 108, \dodoi{10.1093/mnras/staa715}

\bibitem[{Webb {et~al.}(2022)Webb, Gandhi, Brogi, Birkby, de~Mooij, Snellen, \& Zhang}]{Webb_2022}
Webb, R.~K., Gandhi, S., Brogi, M., {et~al.} 2022, \mnras, 514, 4160, \dodoi{10.1093/mnras/stac1512}

\bibitem[{Winn {et~al.}(2008)Winn, Henry, Torres, \& Holman}]{Winn_2008}
Winn, J.~N., Henry, G.~W., Torres, G., \& Holman, M.~J. 2008, \apj, 675, 1531, \dodoi{10.1086/527032}

\bibitem[{Wolf {et~al.}(2007)Wolf, Laughlin, Henry, Fischer, Marcy, Butler, \& Vogt}]{Wolf_2007}
Wolf, A.~S., Laughlin, G., Henry, G.~W., {et~al.} 2007, \apj, 667, 549, \dodoi{10.1086/503354}

\bibitem[{{Wright} {et~al.}(2023){Wright}, {Nugroho}, {Brogi}, {Gibson}, {de Mooij}, {Waldmann}, {Tennyson}, {Kawahara}, {Kuzuhara}, {Hirano}, {Kotani}, {Kawashima}, {Masuda}, {Birkby}, {Watson}, {Tamura}, {Zwintz}, {Harakawa}, {Kudo}, {Hodapp}, {Jacobson}, {Konishi}, {Kurokawa}, {Nishikawa}, {Omiya}, {Serizawa}, {Ueda}, {Vievard}, \& {Yurchenko}}]{Wright2023}
{Wright}, S. O.~M., {Nugroho}, S.~K., {Brogi}, M., {et~al.} 2023, \aj, 166, 41, \dodoi{10.3847/1538-3881/acdb75}

\bibitem[{Yan {et~al.}(2023)Yan, {Nortmann, L.}, {Reiners, A.}, {Piskunov, N.}, {Hatzes, A.}, {Seemann, U.}, {Shulyak, D.}, {Lavail, A.}, {Rains, A. D.}, {Cont, D.}, {Rengel, M.}, {Lesjak, F.}, {Nagel, E.}, {Kochukhov, O.}, {Czesla, S.}, {Boldt-Christmas, L.}, {Heiter, U.}, {Smoker, J. V.}, {Rodler, F.}, {Bristow, P.}, {Dorn, R. J.}, {Jung, Y.}, {Marquart, T.}, \& {Stempels, E.}}]{Yan_2023}
Yan, F., {Nortmann, L.}, {Reiners, A.}, {et~al.} 2023, A\&A, 672, A107, \dodoi{10.1051/0004-6361/202245371}

\bibitem[{Zhang {et~al.}(2018)Zhang, Knutson, Kataria, Schwartz, Cowan, Showman, Burrows, Fortney, Todorov, Desert, Agol, \& Deming}]{Zhang_2018}
Zhang, M., Knutson, H.~A., Kataria, T., {et~al.} 2018, \aj, 155, 83, \dodoi{10.3847/1538-3881/aaa458}

\bibitem[{Öberg {et~al.}(2011)Öberg, Murray-Clay, \& Bergin}]{Oberg_2011}
Öberg, K.~I., Murray-Clay, R., \& Bergin, E.~A. 2011, \apj, 743, L16, \dodoi{10.1088/2041-8205/743/1/l16}

\end{thebibliography}
\bibliographystyle{aasjournal}

\end{document}